\documentclass[onecolumn,aps,floatfix,prd,notitlepage,superscriptaddress,showpacs]{revtex4-1}

\usepackage{latexsym}
\usepackage{amsmath}
\usepackage{amssymb}
\usepackage{latexsym}
\usepackage{xcolor}
\usepackage{graphicx}
\usepackage{cancel}
\usepackage{bbm}
\usepackage{bm}
\usepackage{maybemath}
\usepackage{seceqn}
\usepackage{natbib}
\usepackage{siunitx}

\def\lambdabar{{\mskip2mu\mathchar'26\mkern-9.8mu\lambda}}

\definecolor{kellygreen}{rgb}{0.3, 0.73, 0.09}
\definecolor{duelferred}{rgb}{0.7, 0.2, 0.1}
\definecolor{garrosgreen}{rgb}{0.1, 0.4, 0.1}
\definecolor{cambridgeblue}{rgb}{0.1, 0.3, 1.0}
\definecolor{oxfordblue}{rgb}{0.05, 0.2, 0.7}

\newcommand{\grayone}{black}

\newcommand{\dd}{\mathrm{d}}

\newcommand{\ii}{\mathrm{i}}

\newcommand{\calF}{\mathcal{F}}

\newcommand{\calM}{\mathcal{M}}
\newcommand{\calO}{\mathcal{O}}
\newcommand{\calS}{\mathcal{S}}
\newcommand{\calT}{\mathcal{T}}

\newcommand{\calL}{\mathcal{L}}

\newcommand{\thr}{\mathrm{th}}

\newcommand\myapp{\par
  \setcounter{section}{0}
  \setcounter{subsection}{0}
  \setcounter{figure}{0}
  \setcounter{table}{0}
  \renewcommand\thesection{\Alph{section}}
  \renewcommand\thefigure{\Alph{section}\arabic{figure}}
  \renewcommand\thetable{\Alph{section}\arabic{table}}
}

\begin{document}

\title{Calculation of the Decay Rate of Tachyonic Neutrinos
against Charged--Lepton--Pair and Neutrino--Pair Cerenkov Radiation}

\author{Ulrich D. Jentschura}

\affiliation{Department of Physics,
Missouri University of Science and Technology,
Rolla, Missouri 65409, USA}

\affiliation{MTA--DE Particle Physics Research Group,
P.O.Box 51, H--4001 Debrecen, Hungary}

\author{Istv\'{a}n N\'{a}ndori}

\affiliation{MTA--DE Particle Physics Research Group,
P.O.Box 51, H--4001 Debrecen, Hungary}

\author{Robert Ehrlich}

\affiliation{$^b$Department of Physics, George Mason University, Fairfax, 
Virginia 22030, USA}

\begin{abstract}
We consider in detail the calculation of the 
decay rate of high-energy superluminal neutrinos against
(charged) lepton pair Cerenkov radiation (LPCR), and 
neutrino pair Cerenkov radiation (NPCR), 
i.e., against the 
decay channels $\nu \rightarrow \nu \, e^+ \, e^-$ and
$\nu \rightarrow \nu \, \overline\nu \, \nu$.
Under the hypothesis of a tachyonic nature of neutrinos,
these  decay channels put constraints on the lifetime 
of high-energy neutrinos for terrestrial 
experiments as well as on cosmic scales.
For the oncoming neutrino, we use the 
Lorentz-covariant tachyonic relation 
$E_\nu = \sqrt{\vec p^2 - m_\nu^2}$,
where $m_\nu$ is the tachyonic mass parameter.
We derive both threshold conditions as well as 
decay and energy loss rates, using the plane-wave 
fundamental bispinor solutions of the tachyonic Dirac equation.
Various intricacies of rest frame versus lab frame
calculations are highlighted.
The results are compared to the observations of high-energy
IceCube neutrinos of cosmological origin.
\end{abstract}


\maketitle

\tableofcontents

%
%
\section{Introduction}
\label{sec1}

We describe a calculation of the decay rate and energy loss rate of tachyonic
(superluminal, ``faster-than-light'') neutrinos due to 
(charged) lepton pair Cerenkov radiation (LPCR)
and neutrino pair Cerenkov radiation (NPCR).
These two decay channels proceed via virtual $Z^0$ bosons.
The processes are kinematically allowed for tachyonic (space-like)
neutrinos, and in the case of LPCR,
above a certain energy threshold dependent on the neutrino mass.
We base our treatment on a
Lorentz-covariant theory of tachyonic (faster-than-light) spin-$1/2$ particles,
i.e., on the tachyonic Dirac (not Majorana)
equation~\cite{ChHaKo1985,ChKoPoGa1992,Ko1993,ChKo1994,Ch2012}. 
Solutions of this equation~\cite{JeWu2012epjc,Je2012imag,JeWu2012jpa,%
JeWu2013isrn,JeWu2014}
fulfill the Lorentz covariant dispersion relation 
$E = (\vec k^{\,2} - m_\nu^2)^{1/2}$, 
where $E$ is the energy and $\vec k$ is the spatial momentum vector,
while $m_\nu$ is the tachyonic parameter, corresponding to a
negative Lorentz-invariant mass square $-m_\nu^2$.  The quantity 
$p^\mu \, p_\mu = E^2 - \vec k^{\,2} = -m_\nu^2$ is Lorentz invariant. 
(Again, we shall assume here that neutrinos are Dirac particles 
and use the tachyonic Dirac 
equation~\cite{ChHaKo1985,ChKoPoGa1992,Ko1993,ChKo1994,Ch2012} as a candidate for their
physical description.)

Tachyonic kinematics are somewhat counter-intuitive.
For example, tachyons accelerate as they lose energy.
For a subluminal (``tardyonic'') particle,
one can perform a Lorentz transformation into the 
rest frame where the spatial momentum $k'$ of the 
particle vanishes. For a tachyonic particle,
one can show that, starting from a state with 
real (as opposed to complex) energy $\sqrt{k^2 - m_\nu^2}$,
that the Lorentz-transformed momentum always remains 
greater or equal than than $m_\nu$, i.e., $k' \geq m_\nu$,
and the Lorentz-transformed energy $E'$ remains real~\cite{JeWu2014}.
One thus cannot possibly enter the rest frame where
otherwise we would have $k' = 0$,
and the energy would become complex.
All that we can do for a tachyon is to 
transform into a frame where the Lorentz-transformed energy 
of the neutrino vanishes, i.e., we can enforce $E' =0$, but not $k'=0$.
The latter frame constitutes a (distant)
analogue of the ``rest frame'' of a tachyonic particle,
where according to the classical dispersion
relation, the fact that  $E = m_\nu/\sqrt{v_\nu^2 - 1} = 0$
otherwise implies an infinite velocity $v_\nu = \infty$.
All of these intricacies have to be taken into account
in the calculation of threshold conditions and decay rates.

Here, we analyze the decay of 
energetic tachyonic neutrino via LPCR and NPCR.
In the calculation of the decay and energy loss rates,
we make extensive use of a recently developed formalism 
which expresses the solutions of the tachyonic 
Dirac equation in terms of helicity 
spinors~\cite{JeWu2012epjc,Je2012imag,JeWu2012jpa,JeWu2013isrn,JeWu2014}.
Indeed, helicity remains a good quantum number 
for tachyonic solutions while the chirality operator
does not commute with the tachyonic Dirac Hamiltonian,
a fact which, among other things, leads to a natural explanation for the 
$V-A$ structure of the weak leptonic current~\cite{Ci1998}.
On a different issue, in particle physics, one usually carries out sums over the 
bispinor solutions using Casimir's trick~\cite{Gr1987},
which is based on sum formulas that allow one to 
express the sum over the spin orientations 
of the spin-$1/2$ in a very concise, analytic form.
For the tachyonic Dirac equation, the analogous sum formulas
have recently been found~\cite{JeWu2012epjc,JeWu2013isrn}, in the helicity basis.

A further complication arises because the 
time ordering along a space-like trajectory
of a tachyonic neutrino is not unique.
For a straight space-like trajectory with velocity $v_\nu > c$,
it is possible to boost into a system with 
velocity $u = c^2/v_\nu$, where the tachyonic particle
assumes an infinite velocity, according to the 
velocity addition theorem $v' = (v_\nu - u)/(1 - u\,v_\nu/c^2)$.
Because $v_\nu > c$, we still have $u = c^2/v_\nu < c$,
which makes the boost permissible.
A boost into any frame with velocity
$u'$ (with $u < u' < c$) will reverse the time ordering
along a tachyonic trajectory.
The time ordering problem for a tachyonic trajectory
is connected with the problem that 
some fundamental tachyonic particle operators necessarily transform
into tachyonic antiparticle operators upon 
Lorentz transformation~\cite{BiDeSu1962,ArSu1968,DhSu1968,BiSu1969,%
Fe1967,Fe1978,Re2009,Bi2009,Bo2009}.
For the decay of a tachyonic neutrino into an 
electron-positron pair, this consideration implies
that one is at risk of picking up a contribution
from neutrino-antineutrino annihilation when 
considering the decay of an incoming 
tachyonic neutrino. One can avoid this pitfall
by introducing helicity projectors;
these eliminate the spurious contribution
from the annihilation channel.
A clear exposition of the underlying formalism is one 
of the purposes of the current investigation.

The observation of highly energetic cosmic neutrinos by 
the IceCube collaboration~\cite{AaEtAl2013,AaEtAl2014,Bo2015}
puts constraints on the superluminality of 
neutrinos because they need to ``survive''
the decay processes $\nu \rightarrow \nu \, e^+ \, e^-$ and
$\nu \rightarrow \nu \, \overline \nu \, \nu$.
So, if the decay rate is otherwise sufficiently 
large in order to account for a substantial energy loss
on interstellar time and distance scales,
then one may relate the tachyonic threshold 
to a conceivable high-energy cutoff 
of the cosmic neutrino spectrum at a threshold 
energy $E = E_\thr \approx \SI{2}{PeV}$~\cite{AaEtAl2013,AaEtAl2014,Bo2015}.
Namely, in principle (see Ref.~\cite{JeEh2016advhep}),
the tachyonic theory allows us to express the 
threshold energy as a function of the electron and neutrino
masses, $E_\thr = f(m_e, m_\nu)$; a specific value of the 
threshold thus implies a definite value of $m_\nu$ and 
also determines a numerical value for 
$\delta_\nu = v_\nu^2/c^2 - 1$, because of the 
dispersion relation $E = m_\nu/\delta_\nu^{1/2}$.
However, all these conjectures crucially depend on the 
overall magnitude of the decay and energy loss rates:
If these should turn out to be negligible 
on cosmic distance and time scales, then it will 
be impossible to relate the tachyonic mass parameter to the 
cutoff; hence, it is very important to have explicit
results for the decay rates at hand.

We here continue a series of investigations, continued over the 
last decades, on tachyonic particles~\cite{BiDeSu1962,DhSu1968,BiSu1969,%
Fe1967,Fe1978,ReMi1974,MaRe1980} in general and 
spin-$1/2$ particles and the 
superluminal neutrino hypothesis in 
particular~\cite{ChHaKo1985,Ch2000,Ch2002,Ch2002a,Re2009,Bi2009,Bo2009}.
The latter include Lorentz-violating 
models~\cite{CoGl2011,BiEtAl2011,BeLe2012,DiKoMe2014,Di2014,Ta2014,StEtAl2015,Li2015}
which lead to superluminality;
such models have been applied to the analysis of astrophysical 
data~\cite{StSc2014,St2014}.
Energy loss mechanisms due to LPCR have been subjected to 
alternative statistical analyses~\cite{Ma2014prd},
and compared to other energy loss mechanisms,
e.g., due to neutrino splitting~\cite{MaEtAl2010}.
Neutrino speed modifications have been
linked to conceivable (local) variations in fundamental constants~\cite{FlPo2012},
and a connection of neutrino speed and neutrino 
flavor oscillations has been highlighted in Ref.~\cite{SaVa2013}.
Gravitational interactions have also been linked 
to neutrino speed modifications~\cite{HoAs2012,NoJe2015tach}.
In terms of conceptual questions underlying 
both spinless as well as spin-$1/2$ tachyonic theories, 
including the stability of the vacuum, we refer to 
the discussion in Refs.~\cite{JeWu2013isrn,JeEtAl2014}.
A lengthy further discussion on the conceptual issues underlying the 
tachyonic model would otherwise be beyond the current paper,
which already is quite verbose.

We organize our investigations as follows.
In Sec.~\ref{sec2}, we derive the energy threshold for LPCR 
as a function of the tachyonic mass parameter $m_\nu$.
The derivation of the decay and energy 
loss rates due to lepton pair Cerenkov radiation 
is described in Sec.~\ref{sec3}.
For neutrino pair Cerenkov radiation,
formulas can be found in Sec.~\ref{sec4}.
Phenomenological consequences (IceCube data) are discussed in Sec.~\ref{sec5}.
Units with $\hbar = c = \epsilon_0 = 1$ are used throughout 
this paper.

\begin{figure}[t!]
\begin{center}
\begin{minipage}{0.8\linewidth}
\begin{center}
\includegraphics[width=0.8\linewidth]{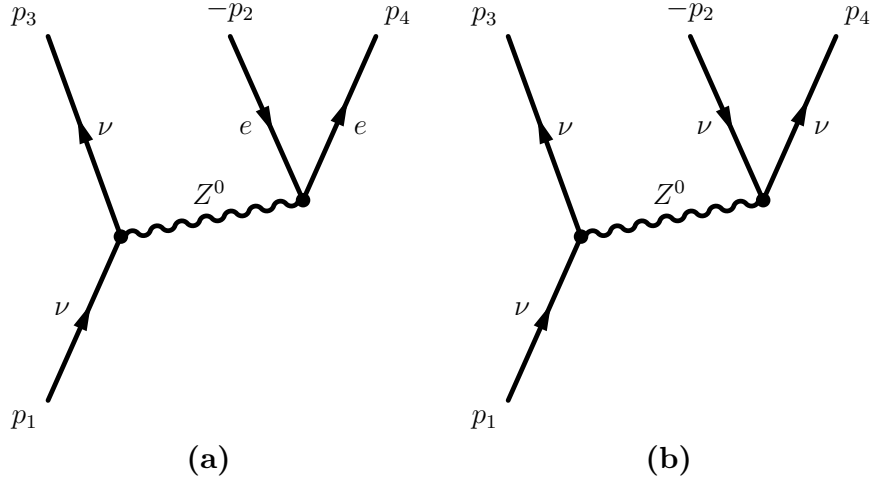}
\end{center}
\caption{\label{fig1} The incoming and outgoing momenta,
for lepton pair Cerenkov radiation (LPCR),
from a tachyonic neutrino, are used as indicated in the 
Feynman diagram (a). The arrow of time is from bottom to top.
The four-momentum of the incoming highly energetic superluminal neutrino 
carries a subscript $1$; it decays into a 
neutrino of lesser energy (subscript $3$), while producing an electron-positron 
pair (subscripts $2$ and $4$). In Fig.~(b), 
the decay products are tachyonic neutrinos of the same mass eigenstate
as the oncoming one. The depicted process is referred to as 
neutrino-pair Cerenkov radiation (NPCR).}
\end{minipage}
\end{center}
\end{figure}

%
%
\section{Thresholds, Fermi Theory and Tachyonic Decays}
\label{sec2}

%
%
\subsection{Tachyonic Lepton Pair Threshold Based on a Space--Like Dispersion Relation}
\label{threshLPCR}

We consider the process shown in Fig.~\ref{fig1}(a),
which is lepton-pair Cerenkov radiation (LPCR).
The threshold condition reads as
$q^2 = (E_3 - E_1)^2 - (\vec k_3 - \vec k_1)^2 \geq 4 m_e^2$,
where $q$ is the four-momentum of the virtual $Z^0$ boson,
while the incoming and outgoing neutrino momenta 
are $p_1^\mu = (E_1, \vec k_1)$ and $p_3^\mu = (E_3, \vec k_3)$.
Threshold is reached when, depending on the geometry, 
the energy transfer from initial to final state is maximum,
while the spatial momentum transfer is minimum.
This implies that a larger spatial momentum transfer actually is
disfavored from a point of view of pair production,
because it leads to lesser values of $q^2$.
In other words, the greater the spatial momentum transfer,
the smaller is the four-momentum transfer.
Geometrically, we want the outgoing spatial momentum to be as
close to the incoming spatial momentum as possible.
At threshold, we can thus safely assume that
the final neutrino state actually propagates into the
same direction as the initial state.

Threshold is reached for a collinear geometry of maximum symmetry.
The incoming and outgoing tachyonic particles
are on the mass shell, i.e., $E_1 = \sqrt{k_1^{\,2} - m_\nu^2}$ and
$E_3 = \sqrt{k_3^{\,2} - m_\nu^2}$.
The four-vector notation can thus be reduced to just two components,
$q = (E_3, k_3) - (E_1, k_1)$,
and the momentum transfer $q^2$ carried by the 
$Z^0$ boson therefore reads as follows,
\begin{equation} 
q^2 = (E_1 - E_3)^2 - (k_1 - k_3)^2 \,.
\end{equation}
Electron-positron pair production threshold is reached at
\begin{equation}
\label{cond1}
q^2 = \left( \sqrt{k_1^{\,2} - m_\nu^2} -
\sqrt{k_3^{\,2} - m_\nu^2} \right)^2 -
(k_1 - k_3)^2 = 4 m_e^2 \,.
\end{equation}
For minimum energy and momentum of the final 
neutrino state, we have $E_3 = 0$ and $k_3 = m_\nu$. Then,
the condition~\eqref{cond1} transforms into
\begin{equation}
\label{k1th}
(k_1)_{\thr} = 2 \frac{m_e^2}{m_\nu} + m_\nu \,.
\end{equation}
The energy of the tachyonic neutrino at threshold is given as
\begin{equation}
\label{E1th}
(E_1)_{\thr} = 
\sqrt{ (k_1)^2_{\thr} - m_\nu^2 } = 
2 \, \frac{m_e}{m_\nu}  \, \sqrt{ m_e^2 + m_\nu^2} 
\approx 
2 \, \frac{m_e^2}{m_\nu} + m_\nu + \calO\left( \frac{m_\nu^3}{m_e^2} \right) \,,
\end{equation}
where the latter approximation is valid
for $\delta \ll 1$. One can rewrite this result,
based on the tachyonic dispersion relation $m_\nu = E_1 \, \sqrt{v_\nu^2 - 1} =
E_1 \, \sqrt{\delta_\nu}$,
\begin{equation}
\label{Eth_form}
(E_1)_{\thr} \approx (k_1)_{\thr} = 2 \frac{m_e^2}{m_\nu} + m_\nu 
\approx 2 \frac{m_e^2}{m_\nu} =
2 \frac{m_e^2}{(E_1)_{\thr} \, \sqrt{\delta_\nu}} 
\qquad
\Rightarrow
\qquad
(E_1)_{\thr} \approx \sqrt{2} \, \frac{m_e}{\delta_\nu^{1/4}} \,.
\end{equation}
We note that the tachyonic threshold is a definite function
of the mass parameters $m_\nu$ and $m_e$ of the 
tachyonic neutrino and of the electron (positron),
respectively.

%
%
\subsection{Tachyonic Neutrino Pair Threshold Based on a Space--Like Dispersion Relation}
\label{threshNPCR}

In the previous section, we found that for electron-positron 
(charged lepton) pair production, 
threshold is reached for a collinear geometry.
In order to investigate the presence or absence of a 
threshold for tachyonic pair production
(i.e, with two outgoing tachyons), it is
instructive to have a look at various geometries.
Let us consider a pair of tachyons, 
both of them on the mass shell, $E^2 - \vec k^{\,2} = -m_\nu^2$.
The outgoing particles of the 
pair are labeled with the indices $2$ and $4$,
as in Fig.~\ref{fig1}.
If we assume that the tachyons are 
emitted collinearly and with the same energy, then
$p^\mu = (E, \vec k) = p_2^\mu = (E_2, \vec k_2) = 
p_4^\mu = (E_4, \vec k_4)$, and
\begin{equation}
q^2 = (p_2 + p_4)^2 = 4 p^\mu p_\mu = 4 \, (\vec k^2 - m_\mu^2) - 4 \, \vec k^{\,2} 
= -4\, m_\mu^2 \,,
\end{equation}
which is negative. For two neutrinos of different energy, 
emitted collinearly. i.e., with $\vec k_2 = k_2 \, \hat{\rm e}_z$
and $\vec k_4 = k_4 \, \hat{\rm e}_z$), one has
\begin{subequations}
\begin{align}
E_2 =& \; \sqrt{ k_2^{\,2} - m_\mu^2} \,,
\qquad
E_4 = \sqrt{ k_4^{\,2} - m_\mu^2} \,,
\\[0.1133ex]
\label{q2plot}
q^2 =& \; 
\left( \sqrt{ k_2^{\,2} - m_\mu^2} +
\sqrt{ k_4^{\,2} - m_\mu^2} \right)^2 - 
( k_2 + k_4 )^2 \,.
\end{align}
\end{subequations}
For small tachyonic mass parameter $m_\nu$,
a Taylor expansion yields
\begin{align}
\label{asymp}
q^2 = -\left( 2 + \frac{k_2}{k_4} + \frac{k_4}{k_2} 
\right) \, m_\nu^2 + {\cal O}(m_\nu^4) \,.
\end{align}
In the limits $k_2 \to 0$, $k_4 \to \infty$ (or vice versa),
$q^2$ assumes very large negative numerical values,
demonstrating the absence of a lower threshold.

One might ask, however, if there is perhaps 
a higher cutoff for the allowed $q^2$ 
in relativistic tachyonic pair production kinematics.
For the production of an anti-collinear pair,
one has
\begin{subequations}
\begin{align}
\vec k_2 =& \; k_2 \, \hat{\rm e}_z \,,
\qquad
\vec k_4 = -k_4 \, \hat{\rm e}_z \,,
\\[0.1133ex]
E_2 =& \; \sqrt{ k_2^{\,2} - m_\mu^2 } \,,
\qquad
E_4 = \sqrt{ k_4^{\,2} - m_\mu^2 } \,,
\\[0.1133ex]
q^2 =& \; 
\left( \sqrt{ k_2^{\,2} - m_\mu^2 } +
\sqrt{ k_4^{\,2} - m_\mu^2} \right)^2 - 
( k_2 - k_4 )^2 
\nonumber\\[0.1133ex]
=& \; 4 k_2 k_4 + {\cal O}(m_\nu^2) \,.
\end{align}
\end{subequations}
For large $k_1$ and $k_2$, this expression assumes arbitrarily large
positive numerical values.
The only condition relevant to 
the allowed range of $q^2$ for tachyonic
pair production thus is 
\begin{equation}
\label{q2range}
-\infty < q^2 < \infty \,,
\qquad
q^0 > 0 \,.
\end{equation}
This result has important consequences for the 
calculation of neutrino-pair Cerenkov radiation 
(see Fig.~\ref{fig1}(b)).

%
%
\subsection{Tachyonic Maximum Momentum Transfer and Fermi Theory}
\label{maximum_momentum}

One crucial question one might ask concerns the 
applicability of Fermi theory for the 
decay processes shown in Figs.~\ref{fig1}(a) and~(b),
in the high-energy regime. 
The question is whether the condition $q^2 \ll M_Z^2$,
which ensures the applicability of Fermi theory,
remains valid for a highly energetic, oncoming neutrino.
Concerning this question, we first recall that,
as already shown, threshold for pair production
is reached for collinear geometry, 
i.e., when the final neutrino momentum $k_3$ is along the same direction
as the initial state momentum $k_1$.
This implies that the maximum momentum transfer,
for given energy $E_1$ of the incoming particle,
also is reached for collinear geometry.
Reducing space to one dimension, we find for the 
square $q^2$ of the momentum transfer,
\begin{align}
q^2 =& \;
\left( \sqrt{ k_1^2 - m_\nu^2 }  - \sqrt{ k_3^2 - m_\nu^2 } \right)^2 - (k_1 - k_3)^2 
\nonumber\\[0.1133ex]
=& \;
2 \, \left( k_1 k_3
- \sqrt{ k_1^2 - m_\nu^2 } \, \sqrt{ k_3^2 - m_\nu^2 } - m_\nu^2 \right) \,.
\end{align}
For given $k_1$,
maximum four-momentum transfer is reached when the momentum of the 
outgoing particle is equal to $k_3 = m_\nu$, and thus $E_3 = 0$.
This implies that
\begin{equation}
\label{q2max}
q^2_{\rm max} = 
2 \, m_\nu \left( k_1 - m_\nu\right) \approx 2 k_1 m_\nu \,.
\end{equation}
The condition for using the effective Fermi theory 
for the virtual $Z^0$ boson exchange in Figs.~\ref{fig1}(a) and~(b) is
$q^2 \ll M_Z^2$, which in the high-energy limit 
can be reformulated as 
\begin{equation}
q^2 \approx 2 k_1 m_\nu \approx
2 E_1 m_\nu \ll M_Z^2  \,,
\qquad
\qquad
E_1 \ll \frac{M_Z^2}{m_\nu} \sim
\frac{(10^{11} \, {\rm eV})^2}{1 \, {\rm eV}} \sim
10^{22} \, {\rm eV} \,.
\end{equation}
where we have conservatively estimated the 
neutrino mass parameter to be on the order of $1 \, {\rm eV}$. 
(In general, one estimates 
the neutrino masses to be of order $(0.01 \div 0.05) \, {\rm eV}$,
see Sect.~1 of Ref.~\cite{StVi2006} 
and the discussion around Eq.~(14.21) of Ref.~\cite{PDG2014}.)
The condition 
$E_1 \ll 10^{22} \, {\rm eV} = 10^7 \, {\rm PeV}$ is easily 
fulfilled by the most energetic neutrinos seen by the
IceCube collaboration~\cite{AaEtAl2013,AaEtAl2014},
which do not exceed $\sim 2 \, {\rm PeV}$ in energy.
Hence, we can safely assume Fermi theory to be valid 
in the entire range of incoming neutrino energies relevant
for the current investigation.

\begin{figure}[t!]
\begin{center}
\begin{minipage}{0.9\linewidth}
\begin{center}
\includegraphics[width=0.91\linewidth]{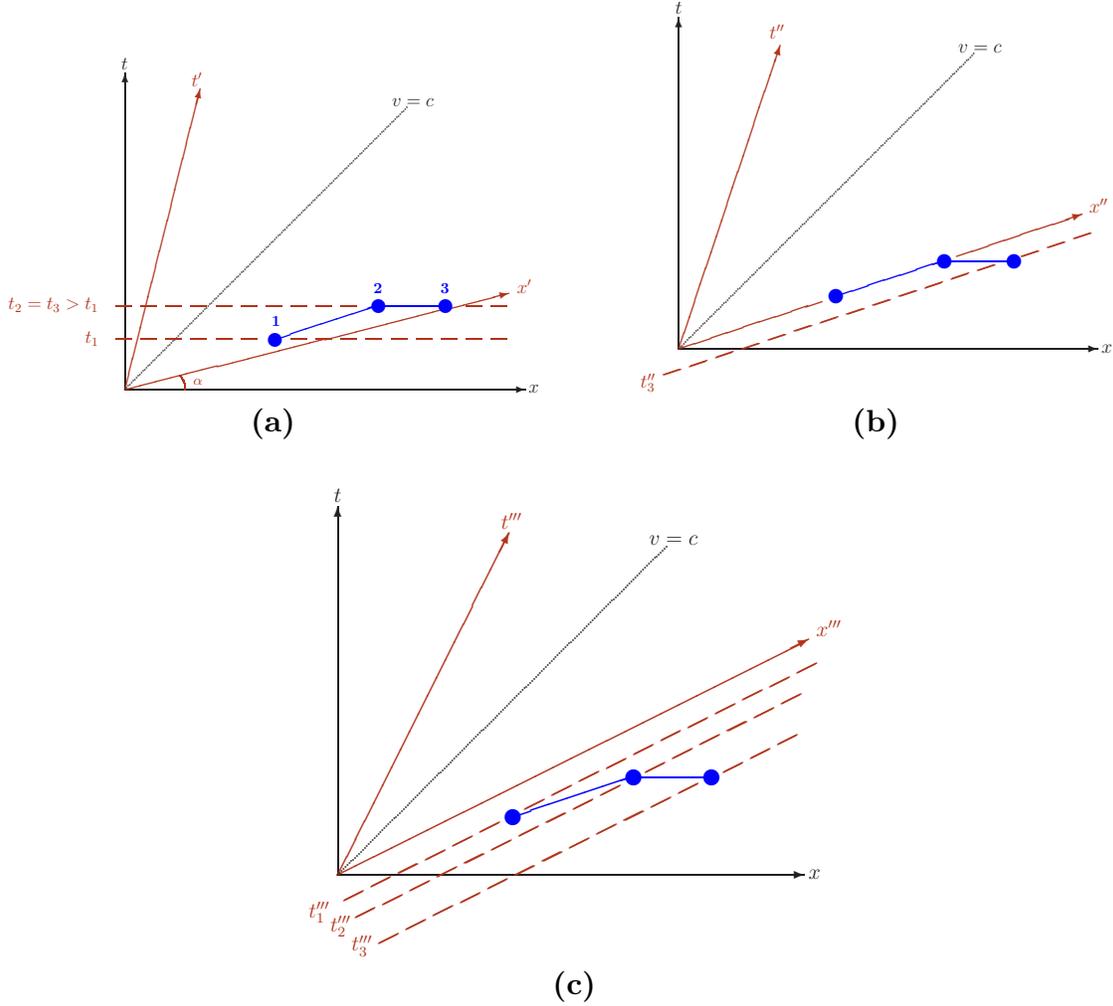}
\end{center}
\caption{\label{fig2}
An incoming tachyonic neutrino follows the world line
$1 \mapsto 2 \mapsto 3$, decaying into a zero-energy,
infinitely fast neutrino [Fig.~(a)].
In the primed frame in Fig.~(a), the time ordering 
of the trajectory $2 \mapsto 3$ is reversed.
The initial neutrino has turned into a 
zero-energy decay ``product'' in Fig.~(b).
Complete reversal of the time ordering of the decay 
process takes place in Fig.~(c), where the 
moving observer (in the triple-primed frame)
interprets the process as the decay of an
incoming antineutrino along the trajectory 
$3 \mapsto 2 \mapsto 1$. Further explanations are in the text.}
\end{minipage}
\end{center}
\end{figure}

\begin{figure}[t!]
\begin{center}
\begin{minipage}{0.7\linewidth}
\begin{center}
\includegraphics[width=0.8\linewidth]{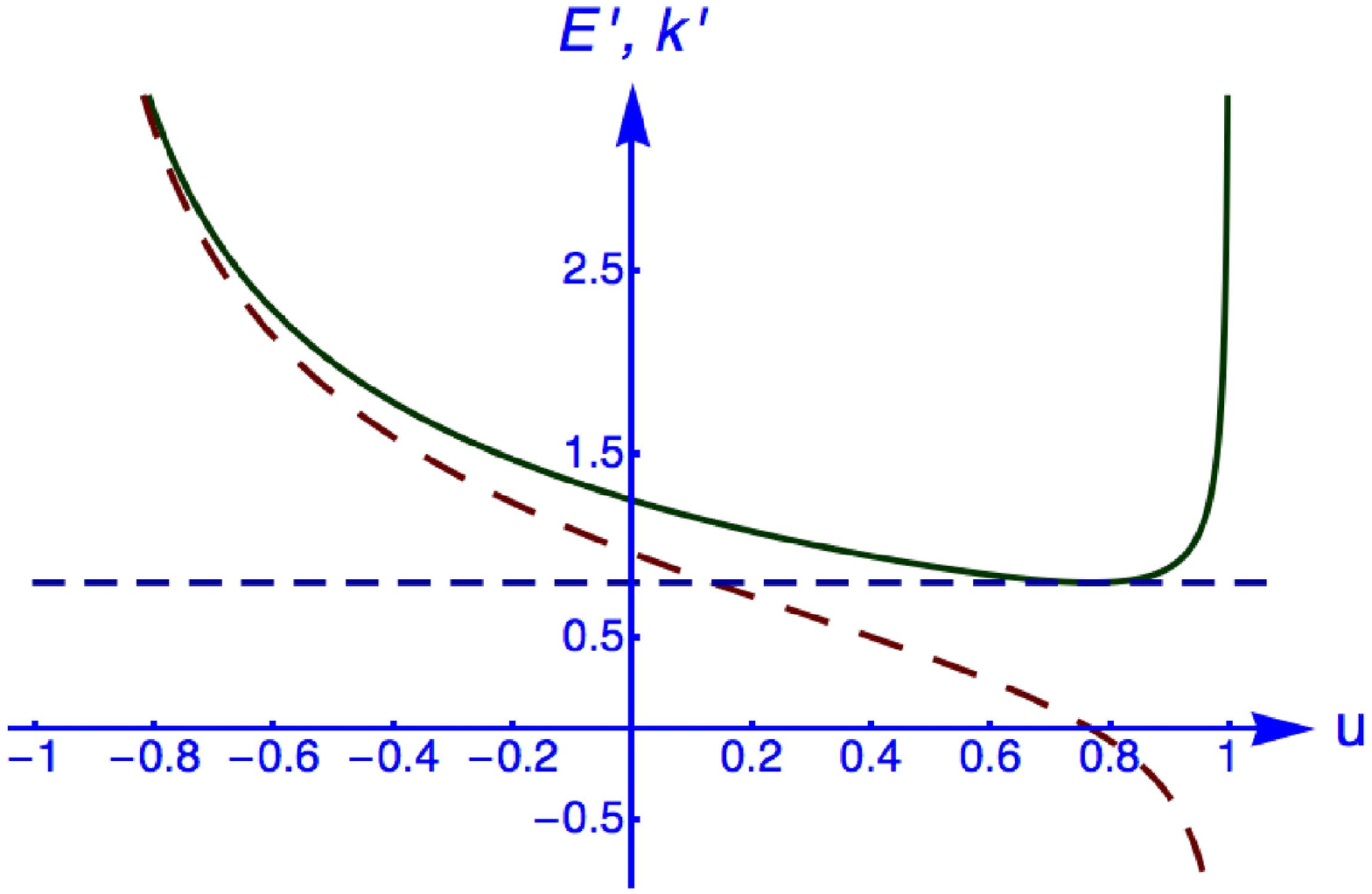}
\end{center}
\caption{\label{fig3} 
Lorentz-transformed momentum vector $k'$ and 
transformed energy $E'$ of a tachyonic 
neutrino, for boost velocities $-1 < u < 1$.
In the lab frame ($u=0$), we have $k = 1.25$ and $-m_\nu^2 = -(0.8)^2$.
Under a Lorentz transformation,
the modulus of the momentum vector $k'$ indeed never gets smaller than
$k' > m_\nu$ (see the solid curve). The energy $E'$ (indicated by long dashes), 
however, can go to zero, 
and in fact changes sign at the point where the modulus 
of the momentum vector just becomes equal to $k' = m_\nu$
which is the point of infinite velocity.
(The constant curve $k' = m_\nu$ is indicated via short dashes.)
When the energy $E'$ changes sign, the propagation direction of the neutrino 
changes sign, and it moves in the negative $x$ direction
as opposed to the positive $x$ direction. 
From the plot (solid curve), one might think that the momentum component
along the boost does not change sign, but this is not physical.
From the Minkowski diagram (see Fig.~\ref{fig2}), one sees that the neutrino 
is still moving along the positive $x$ axis, but with the time ordering
of the start and end point interchanged.}
\end{minipage}
\end{center}
\end{figure}

%
%
\subsection{``Rest'' Frame of the Tachyon}
\label{whywecannot}

Let us briefly analyze the role of the ``rest frame'' of the
faster-than-light, incoming neutrino
in the context of the tachyonic dispersion relation $E = \sqrt{k^2 - m_\nu^2}$.
As is evident from a Minkowski diagram (see Fig.~\ref{fig2}),
the rest frame of the
tachyonic ``space-like'' neutrino cannot be reached via a Lorentz transformation.
By contrast, for a tachyon, it is possible to transform into a frame where the 
{\em time interval} (not the space interval!) swept on the tachyonic trajectory is zero,
i.e., the tachyonic particle assumes an infinite velocity.
This frame of infinite velocity, in some sense,
constitutes the equivalent of the rest frame;
namely, the incoming particle has zero time evolution
(as opposed to zero space evolution),
and thus infinite velocity. According to tachyonic kinematics,
it then has zero energy. For illustration, we consider the boost
into a frame with energy $0 < u < 1$ (see Fig.~\ref{fig3}),
\begin{equation}
\label{lorentz}
E' = \gamma \, (E - u \, k) \,,
\qquad
\qquad
k' = \gamma \, (k - u \, E) \,,
\qquad
\qquad
E = \sqrt{k^2 - m_\nu^2} \,.
\end{equation}
For a boost velocity $u = E/k = \sqrt{k^2 - m_\nu^2}/k < 1$,
we have $E' = 0$, $k' = m$.

One might be tempted to suggest that the decay rate 
calculation could be simplified in the tachyonic ``rest'' frame (with respect to the 
time, not space, i.e., $E' = 0$). 
However, in this frame, one cannot calculate the decay rate.
This is most easily seen from an energy conservation condition.
The oncoming neutrino energy vanishes for infinite velocity ($E'=0$).
Hence, the oncoming particle cannot provide the energy necessary 
to produce an electron-positron pair. 

By contrast and for comparison, for a tardyonic (subluminal) particle, the 
Dirac ``gap'' between positive-energy and negative-energy 
states ensures that the energy of an 
oncoming, say, muon, is always bound by its rest
mass $m_\mu$ from below, even under a Lorentz transformation. 
Hence, the muon decay from rest, with $E = m_\mu \gg 2 m_e$, 
is kinematically possible.
Because the oncoming muon is 
timelike, the emitted virtual $W$ boson can still carry 
enough momentum transfer $q^2 > 0$ in order to produce 
an electron, and an electron antineutrino.
This is not the case for an oncoming tachyonic neutrino,
whose energy is not bound from below, and can in fact vanish.
When the oncoming neutrino energy vanishes, so does the decay rate.

Alternatively, one can observe that in its own rest frame
(the ``real rest frame'' where the tachyon has a vanishing
spatial momentum),
the neutrino has the following properties,
\begin{equation}
v_\nu=0,
\qquad
\qquad
k_\nu = \frac{ m_\nu v_\nu}{\sqrt{v_\nu^2-1}} = 0,
\end{equation}
but
\begin{equation}
E_\nu = m_\nu / \sqrt{0^2 - 1} = \sqrt{k_\nu^2 - m_\nu^2}
= \sqrt{0^2 - m_\nu^2} = \ii \, m_\nu \,.
\end{equation}
The energy becomes imaginary in its own rest frame.
According to Fig.~\ref{fig3}, the rest frame of a space-like,
tachyonic neutrino cannot be reached via a Lorentz transformation,
consistent with the purely real (rather than complex)
quantities which enter Eq.~\eqref{lorentz}.
A further kinematic consideration is illustrative.
Namely, according to Fig.~\ref{fig3},
the energy of the tachyonic particle decreases as one ``chases'' it,
then approaches zero and eventually flips sign
at boost velocity $u$. For boost velocities beyond this point,
the energy becomes negative, or alternatively,
the time ordering of the start and end point of the
trajectory of the tachyon reverses.
A left-handed tachyonic neutrino, for boost velocities beyond $u$,
would be seen as a right-handed antitachyon
moving in the opposite spatial direction,
for the moving observer.
The spatial momentum $k'$, as seen from Fig.~\ref{fig3},
always remains in the region $k' \geq m_\nu$.
The region with imaginary energies
$E_\nu = \pm \ii \, \sqrt{m_\nu^2 - \vec k_\nu^2}$
with $k_\nu < m_\nu$, never can be reached for an
initial plane-wave tachyonic state with
$k_\nu > m_\nu$, via a Lorentz transformation.

These considerations, together, afford an immediate explanation for the 
observation that the final result for the decay rate 
must necessarily vanish with the energy of the oncoming
neutrino, and in fact, shall later be seen to 
contain the neutrino energy as a linear term.
The calculation of the decay rate of the tachyonic 
neutrino needs to be done in the laboratory frame (lab frame).

%
%
\subsection{Particle--Antiparticle Transformations and Tachyonic Decays}
\label{antipart}

A few final considerations regarding the 
time ordering of tachyonic world lines and 
the calculation of the decay rate are in order.
As already emphasized, decay rates are normally calculated most easily in the 
rest frame of the decaying particle.
For tachyons, we cannot go into the true rest frame 
of the decaying particle, 
because the frame with $k' = 0$ cannot be reached
for a tachyon, as already described.
In the case of a tardyonic particle, there is an energy 
gap of twice the rest mass between the 
spectrum of positive-energy (particle) versus negative-energy 
(anti-particle) states.
This energy gap vanishes for tachyons.
A tardyonic oncoming particle state 
cannot transform into an incoming anti-particle state,
irrespective of the Lorentz frame in which the 
process is observed.
This implies that, e.g., for the decay of an oncoming muon into 
a muon neutrino, electron and electron anti-neutrino,
there is no Lorentz frame in which the 
same process would be observed as a time-reversed 
process, i.e., the 
annihilation of an incoming muon antineutrino 
and an incoming muon, into a $W$ boson, 
and the eventual production of an electron and an
electron antineutrino.

Furthermore, it is known that the 
energy of a tachyonic particle may 
change sign upon a Lorentz transformation
(see Fig.~\ref{fig3}),
so that particle trajectories may become anti-particle
trajectories (with the time ordering of 
start and end points reversed).  Indeed, the fact that 
some particle creation and annihilation 
operators transform into anti-particle operator
upon a Lorentz transformation, 
has been mentioned as an important 
problematic aspect of tachyonic field 
theories~\cite{BiDeSu1962,ArSu1968,DhSu1968,BiSu1969,%
Fe1967,Fe1978,Re2009,Bi2009,Bo2009},
while possible re-interpretations
have recently been discussed in Ref.~\cite{JeWu2012epjc}.

For a tachyonic decay of an oncoming initial tachyonic 
neutrino into an electron-positron pair, 
and an energetically lower neutrino, 
this means the following.
The interpretation of the process may depend on the Lorentz frame
in which it is observed; tachyonic trajectories 
have no definite time ordering.
(The only ordering in the tachyonic case concerns the 
helicity: A left-handed particle state will transform
into a right-handed anti-particle state, and 
vice versa.)
The decay of a highly energetic oncoming neutrino
(``Big Bird'', see Refs.~\cite{AaEtAl2013, AaEtAl2014}) 
into a energetically lesser one
(``Tweety'') via electron-positron
pair production is interpreted equivalently as the 
annihilation of an incoming tachyonic
neutrino, and an incoming tachyonic anti-neutrino,
in specific Lorentz frames [see Fig.~\ref{fig2}(b)].
In other Lorentz frames,
it is even reinterpreted as the decay 
of an incoming highly energetic anti-neutrino, 
into a less energetic  anti-neutrino and 
an electron-positron pair [see Fig.~\ref{fig2}(c)].

We now consider the kinematics of the tachyonic 
decays displayed in Fig.~\ref{fig2} in detail.
In Fig.~\ref{fig2}(a), the world-line trajectories of the oncoming 
neutrino (joining space-time points labeled $1$ and $2$)
and of the final zero-energy neutrino 
(joining space-time points labeled $2$ and $3$) are displayed.
When ``chasing'' the decaying neutrino with a Lorentz
boost, transforming the $x$ and $t$ axes 
into $x'$ and $t'$, respectively, 
then from visual inspection, it is evident that 
the time ordering on the 
final decay product trajectory has reversed
($t_2 = t_3$ but $t'_3 < t'_2$).
The decay product has turned into an incoming antineutrino,
and the Lorentz-transformed process describes
neutrino-antineutrino annihilation (into an electron-positron
pair, but the world lines of the decay products are not displayed 
in Fig.~\ref{fig2}). Physical reality has to be
ascribed to both interpretations~\cite{BiDeSu1962,ArSu1968,DhSu1968,BiSu1969}.
The observation of the moving (``primed'') observer is equally valid.
For the lab frame, this means that unless we have a counter-propagating
beam of antineutrinos, the 
neutrino-antineutrino annihilation process does not contribute 
to the discussion of the ``decay'', which only converts
the oncoming highly energetic neutrino into one with lesser energy.

Let us now consider Fig.~\ref{fig2}(b).
The incoming neutrino is chased by a
``faster'' Lorentz boost. The transformed 
axes become $x''$ and $t''$,
and the first, the decaying neutrino,
now constitutes a zero-energy decay 
product, for the decay of an incoming 
antineutrino (time-ordered trajectory $3 \mapsto 2$).
As the boost velocity crosses the 
$x''$ and $t''$ axes, the decay ``product'' (from the 
point of view of the lab frame) has turned
into an incoming, highly energetic, antineutrino, which
in the triple-primed frame in Fig.~\ref{fig2}(c),
decays into an energetically lower antineutrino
(from the point of view of the boosted frame).
At the point where the $x''$ and $t''$ axes are crossed,
the initial, incoming neutrino has transformed 
into an outgoing zero-energy neutrino or anti-neutrino 
state (the interpretation changes exactly at the 
point where the energy changes sign).

What do these considerations imply for the 
description of the tachyonic decay of a neutrino?
We are working in the lab frame, and we 
need to calculate the process in the lab frame.
Processes with incoming anti-neutrinos 
must be excluded from the integration,
because they cannot contribute to the decay of 
an incoming neutrino.
The interpretation of a process
involving tachyons may depend on the Lorentz frame;
for the calculation of the decay rate,
only processes with incoming and 
outgoing positive-energy neutrino states may be considered,
even if these states may transform into 
anti-particle states upon a Lorentz transformation.
The final results are still Lorentz-invariant,
as discussed below in Sec.~\ref{LIV}.

%
%
\section{Lepton Pair Cerenkov Radiation}
\label{sec3}

\color{black}

%
%
\subsection{Interaction Terms in Glashow--Weinberg--Salam Theory}
\label{int_terms}

In order to proceed to the calculation of the decay rate
of the tachyonic, incoming neutrino, 
we briefly compile known Lagrangians from standard electroweak theory
(see also Appendix~\ref{appa}).
We denote the weak coupling constant as $g_w$. 
According to Chap.~12 of Ref.~\cite{ItZu1980},
quantum electrodynamics 
(QED) is described by the coupling of the electron to the photon,
\begin{equation}
\label{L1}
\calL_1 = -g_w \, \sin\theta_W \, 
\left( \overline e \, \gamma^\mu e \right) \, A_\mu \,,
\end{equation}
where $\theta_W$ is the Weinberg angle
and $e$ and $\overline e$ describe the electron-positron 
field operators, while $A_\mu$ is the electromagnetic field operator.
Furthermore, the charged vector boson $W^\pm$ interacts with a neutrino-electron 
current,
\begin{equation}
\label{L2}
\calL_2 = \left\{ -\frac{g_w}{2 \sqrt{2}} \,
\left[ \overline e \, \gamma^\rho \, 
\left( 1 - \gamma^5 \right) \, \nu_e \right] \, W^+_\rho+ {\rm h.c.}
\right\}
+ \left\{ -\frac{g_w}{2 \sqrt{2}} \,
\left[ \overline \nu_\mu \, \gamma^\rho \, 
\left( 1 - \gamma^5 \right) \, \mu \right] \, W^-_\rho + {\rm h.c.}
\right\} \,,
\end{equation}
where the addition of the Hermitian adjoint is necessary 
in order to include the $W^+$ boson.
For the calculation of the muon decay, one needs the full Lagrangian
given in Eq.~\eqref{L2}, even twice, namely,
once for the muon--muon-neutrino current, and a second time 
for the decay of the $W$ into the electron and 
electron antineutrino, i.e., the same current is used in the 
electron and in the neutrino sector.

For the decay process of the tachyonic neutrino, 
one needs the coupling of the neutrino to the 
$Z^0$ boson, 
\begin{equation}
\label{L3}
\calL_3 = - \frac{g_w}{4 \, \cos\theta_W} \, 
\left[ \overline \nu \, 
\gamma^\mu (1 - \gamma^5) \, \nu \right] \, Z_\mu  \,,
\end{equation}
as well as the coupling of the left- and right-handed electron 
to the $Z^0$ boson,
\begin{align}
\label{L4}
\calL_4 =& \; \frac{g_w}{4 \, \cos\theta_W} \,
\overline e \, \left[ \gamma^\mu (1 - \gamma^5) - 
4 \, \sin^2(\theta_W) \, \gamma^\mu \right] \, e \, Z_\mu
\nonumber\\[0.1133ex]
=& \; - \frac{g_w}{2 \, \cos\theta_W} \,
\overline e \, \left[
c_V \, \gamma^\mu - c_A \, \gamma^\mu \, \gamma^5 \right] \, e \, Z_\mu \,,
\nonumber\\[0.1133ex]
c_V =& \; -\frac12 + 2 \, \sin^2(\theta_W) \,,
\qquad
c_A = -\frac12 \,.
\end{align}
The latter form allows us to identify the 
vector-coupling and axial-vector coupling coefficient $c_V$ and $c_A$.
According to Eq.~(12.237) of Ref.~\cite{ItZu1980}, 
the vacuum-expectation value $v$ of the Higgs, 
the weak coupling constants $g_w$ and $g'_w$,
and the masses of the vector gauge bosons $W^\pm$, $Z^0$ and $A$, are related by
\begin{equation}
\label{MWMZ}
M_W = \frac12 \, v \, g_w \,,
\qquad
M_Z = \frac12 \, v \, ( g_w^2 + g_w'^2)^{1/2} \,,
\qquad
M_A = 0 \,,
\qquad
\frac{M_W}{M_Z} 
= \frac{g_w}{( g_w^2 + g_w'^2)^{1/2}} 
= \cos \theta_W \approx \frac{\sqrt{3}}{2} \approx 0.877 \,.
\end{equation}
These values match the experimental observations of 
$M_W = 80.385(15) \, {\rm GeV}/{c^2}$
and $M_Z = 91.1876(21) \, {\rm GeV}/{c^2}$.
The matching with Fermi's effective coupling constant is given as
\begin{equation}
\label{matching}
\frac{G_F}{\sqrt{2}} = \frac{g_w^2}{8 \, M_W^2} \,.
\end{equation}

Let us anticipate a certain consideration regarding the 
prefactors encountered in the calculation of 
invariant matrix elements, in the weak decay of the muon,
and in the weak decay of a tachyonic neutrino.
For the weak decay of the muon, one uses the Lagrangian~\eqref{L2},
whose prefactors give a numerical factor $1/(2 \sqrt{2})^2 = 1/8$.
For the weak decay by LPCR,
we need to use the Lagrangians~\eqref{L3} and~\eqref{L4},
whose combination results in a prefactor
\begin{equation}
\label{pref1}
\left( -\frac{1}{4 \cos\theta_W} \right) \times
\left( -\frac{1}{2 \cos\theta_W} \right) = 
\frac{1}{8 \, \cos^2 \theta_W} \,.
\end{equation}
However, the weak decay of the tachyonic neutrino is mediated by 
a $Z$ boson as opposed to a $W$ boson, which results in a factor
\begin{equation}
\label{pref2}
\frac{1}{8 \, \cos^2 \theta_W} \, \frac{g_w^2}{M_Z^2} =
\frac{g_w^2}{8 \, ( \cos \theta_W M_Z )^2} =
\frac{G_F}{\sqrt{2}} \,,
\end{equation}
which is the same prefactor that we encounter in the invariant 
matrix element for the weak decay of the muon.

%
%
\subsection{Degrees of Freedom in Three--Body Decay}
\label{degrees}

Let us analyze the degrees of freedom in the phase-space 
of the final state, in three-body decay
of a tachyonic neutrino into a less energetic neutrino,
and a light fermion-antifermion pair.
The momentum transfer is 
$q^2 > (2 m_e)^2$ from the first fermion line.
The decay rate is then obtained as an integral over the 
differential decay rate,
\begin{equation}
\dd \Gamma = \dd^3 k_1 \, \dd^3 k_2 \, \dd^3 k_3 \,
\delta^{(4)}(p_1 + p_2 + p_3) \,.
\end{equation}
This decay rate is 9 times differential,
with 4 conservation conditions.
We thus have 5 effective free variables.

These can be assigned as follows: For the decay of a tachyonic neutrino via
pair production, we may fix the three momentum components of the outgoing
neutrino.  Because both the incoming as well as the outgoing neutrino have to
be on the mass shell, this fixes the four-vector $q^\mu = p_1^\mu - p_3^\mu = 
(q^0, \vec q)$
completely. We can then go into the rest frame of the virtual 
$Z^0$ boson and argue that the decay  must be completely symmetric there;
i.e., the electron and positron should come out in directions
exactly opposite of each other. This gives us two more degrees of freedom,
namely, the polar and azimuthal angles of one of the outgoing fermions.

The three momenta of the outgoing neutrino and the two light fermion
angles add up to the five effective degrees of freedom.
So, once we have $q^\mu = (q^0, \vec q)$, we have only two degrees of freedom left
for the electron-positron pair.

\color{\grayone}

%
%
\subsection{Rationale of the Investigation}
\label{rationale}

The rationale behind the calculations reported below
can be summarized as follows. 
We shall approach the eventual calculation of the decay rate of a 
tachyonic neutrino due to electron-positron pair production in 
two steps.
\begin{itemize}
\item Step 1 (Complexities in the lab frame): 
As already emphasized, the tachyonic calculation,
in which we are eventually interested,
requires us to consider amplitudes in the lab frame,
as opposed to the rest frame of the decaying particle.
We thus need experience with calculations in the lab frame.
The calculation of the  muon decay rate is in principle
very well known for the rest frame of the 
decaying particle.  Here, we generalize the calculation to a muon decay
rate calculation in the lab frame, where as we shall see,
the allowed $\vec k_3$ momenta (in the conventions used
for Fig.~\ref{fig1}) are inside an ellipsoid. Lorentz invariance
of the integral over the allowed outgoing momenta
is explicitly shown. 
\item Step 2 (Decay of tachyonic, space-like particles): 
In the calculation 
of the decay rate of the tachyonic neutrino, we
assume (in the spirit of Fig.~\ref{fig1}) that both the 
incoming as well as the outgoing neutrinos are
on the tachyonic mass shell, 
$E_1 = (\vec k_1^{\,2} - m_\nu^2)^{1/2}$ and 
$E_3 = (\vec k_3^{\,2} - m_\nu^2)^{1/2}$.
Under these circumstances, 
tachyonic decay is made possible exclusively due to the mass terms
in the dispersion relations; hence we cannot ignore these 
terms. Furthermore, as we have already discussed,
we need to remember that the region with 
$|\vec k_3| < m_\nu$ actually is excluded from the 
region of allowed tachyonic momenta.
We find that the physically allowed outgoing momenta
are located inside the rotationally symmetric,
shallow hull of a cupola-like 
structure, centered about the axis of the oncoming 
(decaying) neutrino (which we choose to be the 
positive $z$ axis). 
As already anticipated in Sec.~\ref{antipart},
we shall need to explicitly exclude 
from the calculation any 
processes related to neutrino-antineutrino annihilation.
This necessity, in turn, makes the use of the explicit spinor 
solutions of the tachyonic Dirac equation~\cite{JeWu2012epjc,JeWu2013isrn}
necessary.
\end{itemize}
In the calculation, we also need to overcome the pitfall connected with the 
time ordering of the tachyonic trajectories, 
anticipated in Sec.~\ref{antipart}.

\begin{figure}[t!]
\begin{center}
\begin{minipage}{0.7\linewidth}
\begin{center}
\includegraphics[width=0.4\linewidth]{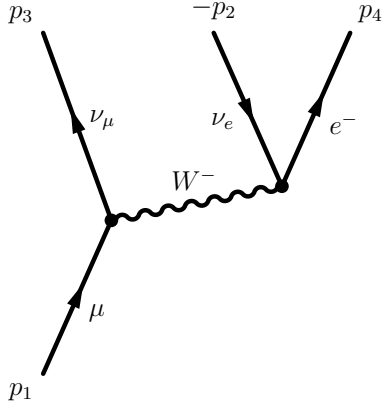}
\end{center}
\caption{\label{fig4} Conventions for muon decay.}
\end{minipage}
\end{center}
\end{figure}

%
%
\subsection{Step 1: Integrating the Muon Decay Width in the Lab Frame}
\label{step1}

We shall consider the weak decay of the muon,
in the conventions of Fig.~\ref{fig4}.
The interaction terms from the electroweak standard model
is used according to Eq.~\eqref{L2}.
For momentum transfers $q^2 \ll M_W^2$, the 
effective four-fermion Lagrangian thus is
\begin{align}
\calL =& \; \frac{g_w^2}{8 \, M_W^2} 
\left( \overline \nu_\mu \, \gamma_\lambda \,
( 1- \gamma^5) \mu \right) \,
\left( \overline e \, \gamma^\lambda \,
( 1- \gamma^5) \nu_e \right) 
\nonumber\\[0.1133ex]
=& \; \frac{G_F}{\sqrt{2}} \,
\left( \overline \nu_\mu \, \gamma_\lambda \,
( 1- \gamma^5) \mu \right) \,
\left( \overline e \, \gamma^\lambda \,
( 1- \gamma^5) \nu_e \right) \,,
\end{align}
where we use the matching~\eqref{matching}.
The Lorentz-invariant matrix element thus is
(within the conventions of Fig.~\ref{fig4})
\begin{equation}
\calM = \frac{G_F}{\sqrt{2}} \,
\left( \overline u(p_3) \, \gamma_\lambda \,
( 1- \gamma^5) u(p_1) \right) \,
\left( \overline u(p_4) \, \gamma^\lambda \,
( 1- \gamma^5) v(p_2) \right) \,.
\end{equation}
Summing over the final spin states and averaging 
over the spin projections of the initial state leads to
\begin{align}
\frac12 \, \sum_{\rm spins} | \calM |^2 =& \; \frac{G_F^2}{4} \,
{\rm Tr} \left[ \cancel{p}_3 \, \gamma_\lambda \,
( 1- \gamma^5) (\cancel{p}_1 + m_\mu ) 
\gamma_\nu  \, ( 1- \gamma^5) \right] \,
{\rm Tr} \left[ \cancel{p}_2 \, \gamma^\lambda \,
( 1- \gamma^5) (\cancel{p}_4 + m_e ) 
\gamma^\nu  \, ( 1- \gamma^5) \right] 
\nonumber\\[0.133ex]
=& \; \frac{G_F^2}{4} \, \left[ 256 \, (p_1 \cdot p_2) \, (p_3 \cdot p_4) \right] 
= 64 \, G_F^2 \, (p_1 \cdot p_2) \, (p_3 \cdot p_4) \,.
\end{align}
The procedure of integration in the rest frame of the 
decaying particle is discussed in Eq.~(10.16) ff.~of Ref.~\cite{Gr1987}.
Furthermore, a mixed approach, where certain intermediate integrals 
are carried out covariantly, and only the final stages
of the calculation are carried out in the rest frame of the decaying
particle, is outlined in Chap.~7.2.2 of Ref.~\cite{LaPa2011}.
In the actual evaluation, in the conventions of 
Fig.~\ref{fig4}, we keep the outgoing 
neutrino momentum as our final integration variable and write
the decay rate in the lab frame as follows
[for the expression in the lab frame, see Ref.~\cite{Jo2011}],
\begin{align}
\label{GammaBasic}
\Gamma =& \; \frac{1}{2 E_1} \,
\int \frac{\dd^3 k_3}{(2 \pi)^3 \, 2 E_3}
\, \left(
\int \frac{\dd^3 k_2}{(2 \pi)^3 \, 2 E_2}
\int \frac{\dd^3 k_4}{(2 \pi)^3 \, 2 E_4}
(2 \pi)^4 \, \delta^{(4)}( p_1 - p_3 - p_2 - p_4 ) \,
\left[ \frac12 \, \sum_{\rm spins} | \calM |^2 \right] \right)
\nonumber\\[0.1133ex]
=& \; \frac{1}{2 E_1} \,
\int \frac{\dd^3 k_3}{(2 \pi)^3 \, 2 E_3}
\, \left(
\int \frac{\dd^3 k_2}{(2 \pi)^3 \, 2 E_2}
\int \frac{\dd^3 k_4}{(2 \pi)^3 \, 2 E_4}
(2 \pi)^4 \, \delta^{(4)}( p_1 - p_3 - p_2 - p_4 ) \,
\left[ 64 G_F^2 ( p_1 \cdot p_2 ) \, (p_3 \cdot p_4) \right] \right)
\nonumber\\[0.1133ex]
=& \; \frac{2 \, G_F^2}{\pi^5 \, (2 E_1)} \,
\int \frac{\dd^3 k_3}{2 E_3}
\, \left(
\int \frac{\dd^3 k_2}{2 E_2}
\int \frac{\dd^3 k_4}{2 E_4}
(2 \pi)^4 \, \delta^{(4)}( p_1 - p_3 - p_2 - p_4 ) \,
( p^\lambda_1 \cdot p_{2\lambda} ) \, (p_{3\rho} \cdot p^\rho_4) \right)
\nonumber\\[0.1133ex]
=& \; \frac{2 \, G_F^2}{\pi^5 \, (2 E_1)} \,
\int \frac{\dd^3 p_3}{2 E_3}
\, \left( p^\lambda_1 p^\rho_3 \, J_{\lambda \rho}(p_1 - p_3) \right)
= \frac{G_F^2}{12 \, \pi^4 \, (2 E_1)} \,
\int \frac{\dd^3 p_3}{2 E_3} \, p^\lambda_1 p^\rho_3 \,
\left( g_{\lambda\rho} \, q^2 + 2 \, q_\lambda \, q_\rho \right)
\nonumber\\[0.1133ex]
=& \; \frac{G_F^2}{12 \, \pi^4 \, (2 E_1)} \, \int \frac{\dd^3 k_3}{2 E_3} \,
\left( p_1 \cdot p_3  \, q^2 + 2 \, ( p_1 \cdot q ) \, (p_3 \cdot q) \right) \,.
\end{align}
We have used the following result,
derived in Eq.~\eqref{resIJK}, which is obtained for 
two outgoing particles with labels $2$ and $4$ which are
on the electronic mass shell $p_2^2 = p_4^2 = m_e^2$,
\begin{align}
J_{\lambda\rho}(q) =& \;
\int \frac{\dd^3 k_2}{2 E_2}
\int \frac{\dd^3 k_4}{ 2 E_4}
\delta^{(4)}( q - p_2 - p_4 ) \,
\left( p_{2 \lambda} \; p_{4 \rho} \right)
\nonumber\\[0.1133ex]
=& \; \sqrt{1 - \frac{4 \, m_e^2}{q^2}} \,
\left[ g_{\lambda \rho} \, \frac{\pi}{24} \,
\left( q^2 - 4 m_e^2 \right) +
q_\lambda \, q_\rho \,
\frac{\pi}{12} \,
\left( 1 + \frac{2 m_e^2}{q^2} \right) \right] 
\nonumber\\[0.1133ex]
\approx & \; 
\frac{\pi}{24} \, g_{\lambda \rho} \, q^2 +
\frac{\pi}{12} \, q_\lambda \, q_\rho \,,
\qquad
\qquad
m_e \to 0 \,.
\end{align}
By symmetry, we have $J_{\lambda\rho}(q) = J_{\lambda\rho}(q)$.
We have carried out the $p_2$ and $p_4$ integrals covariantly.
Then, for the remaining integral over $p_3$, we need the 
appropriate integration limits.
We thus need to integrate
\begin{equation}
\label{GammaInt}
\Gamma = \frac{G_F^2}{12 \, \pi^4 \, (2 E_1)} \, \int \frac{\dd^3 k_3}{2 E_3} \,
\left( p_1 \cdot p_3  \, q^2 + 2 \, ( p_1 \cdot q ) \, (p_3 \cdot q) \right) \,,
\end{equation}
assuming an incoming muon with energy $E_1$ in the positive 
$z$ direction, with the incoming $p_1$ on the muon mass shell,
$p_1^2 = m_\mu^2$. The final $p_3$ describes the muon
neutrino, so that within our approximations $(p_3)^2 \approx 0$.
The domain of the $\dd^3 p_3$ integration in Eq.~\eqref{GammaInt}
contains all four-momenta $p_3$ for which $q^2 = (p_1 - p_2)^2 >0$.

\begin{figure}[t!]
\begin{center}
\begin{minipage}{0.7\linewidth}
\begin{center}
\includegraphics[width=0.5\linewidth]{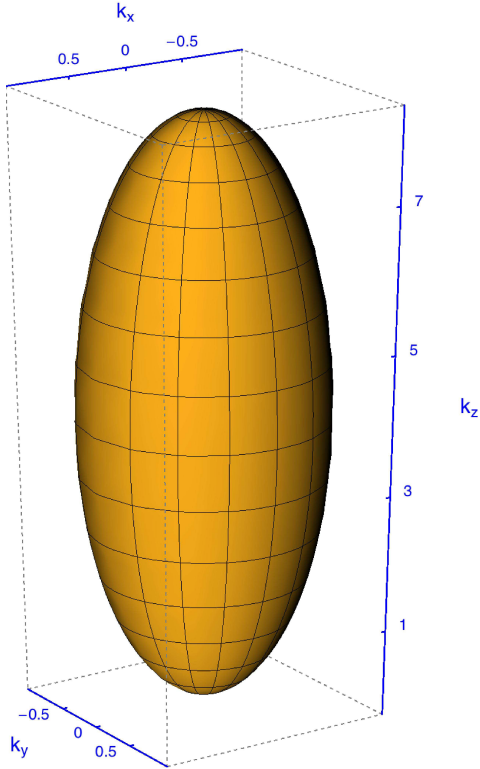}
\end{center}
\caption{\label{fig5} A muon with
wave vector $k_1 = 8$ and mass $m_\mu = 2$ is incoming
along the positive $z$ direction. The
electron mass as well as the neutrino masses
are assumed to be negligible,
and the threshold condition for weak decay into a
muon neutrino, electron and electron anti-neutrino
therefore simplifies to
$q^2 = (p_1 - p_3)^2 > 0$. We investigate the 
boundaries of the volume of allowed $k_3$ vectors,
with Cartesian components 
$k_x = k_{3x}$, $k_y = k_{3y}$, and $k_z = k_{3z}$.
The $z$ components of the allowed
$k_3$ vectors range from $k_{3, {\rm min}} = -0.1231$ to
$k_{3, {\rm max}} = 8.1231$. The geometry of allowed
$k_3$ vectors is that of an ellipsoid, 
rotationally symmetric about the $z$ axis, 
with parameters $a=1$ and $b=4.1231$ as
given in Eq.~\eqref{ellipseparam}.}
\end{minipage}
\end{center}
\end{figure}

In the rest frame of the decaying muon, 
the integration domain would consist of a sphere 
composed of vectors $p_3 = (|\vec k_3|, \vec k_3)$,
with $p_1 = (E_1, \vec 0)$ and $|\vec k_3| \leq  m_\nu/2$.
By contrast, in the lab fame, we consider a muon moving up the $z$ axis, with 
energy $E_1$ and wave vector $\vec k_1$, and an outgoing 
muon neutrino with energy $E_3$ and  wave vector $\vec k_3$,
\begin{equation}
\vec k_1 = k_1 \, \hat e_z \,,
\qquad
E_1 = \sqrt{k_1^2 + m_\mu^2} \,,
\qquad
\vec k_3 = k_\rho \, \hat e_\rho + k_z \hat e_z \,,
\qquad
E_3 = | \vec k_3 | = \sqrt{k_\rho^2 + k_z^2} \,.
\end{equation}
The momentum transfer reads as follows,
\begin{equation}
q^2 = 2 k_1 \, k_z + m_\mu^2 - 2 \sqrt{k_\rho^2 + k_z^2} \,
\sqrt{k_1^2 + m_\mu^2} \,.
\end{equation}
The allowed vectors $\vec k_3$ are located 
inside a rotationally symmetric ellipsoid
(see Fig.~\ref{fig5}), which is centered at the point
$(0,0,k_{z0})$ on the $z$ axis.
Let us denote by $a$ the half axis of the ellipsoid in the 
radial direction (``away'' from the $z$ axis) and by 
$b$ the half axis of the ellipsoid in the $z$ direction (see Fig.~\ref{fig5}).
These half axes are given as follows,
\begin{equation}
\label{ellipseparam}
a = \frac{m_\mu}{2} \,,
\qquad
b = \frac12 \, \sqrt{k_1^2 + m_\mu^2} \,,
\qquad
k_{z0} = \frac{k_1}{2} \,.
\end{equation}
For the integration of the final phase-space in the expression~\eqref{GammaInt},
we need to calculate
\begin{equation}
\Gamma = \frac{G_F^2}{12 \, \pi^4 \, (2 E_1)} \, 
\int\limits_{q^2 = (p_1 - p_3)^2 > 0} \frac{\dd^3 p_3}{2 E_3} \,
\left( p_1 \cdot p_3  \, q^2 + 2 \, ( p_1 \cdot q ) \, (p_3 \cdot q) \right) \,.
\end{equation}
One uses the following parameterization
($k_x \equiv k_{3x}$, $k_y \equiv k_{3y}$, and
$k_z \equiv k_{3z}$),
\begin{align}
k_x = a \, \xi \, \sin \theta \, \cos\varphi \,,
\qquad
k_y = a \, \xi \, \sin \theta \, \sin\varphi \,,
\qquad
k_z = k_{z0} + b \, \xi \, \cos \theta \,.
\end{align}
The Jacobian is 
\begin{equation}
\dd^3 k_3 = \left| \det \left( \begin{array}{ccc}
\dfrac{\partial k_x}{\partial \xi} &
\dfrac{\partial k_x}{\partial \theta} &
\dfrac{\partial k_x}{\partial \varphi} \\[2.133ex]
\dfrac{\partial k_y}{\partial \xi} &
\dfrac{\partial k_y}{\partial \theta} &
\dfrac{\partial k_y}{\partial \varphi} \\[2.133ex]
\dfrac{\partial k_z}{\partial \xi} &
\dfrac{\partial k_z}{\partial \theta} &
\dfrac{\partial k_z}{\partial \varphi}
\end{array} \right) \right|
= a^2 \, b \, \xi \, \sin\theta \,.
\end{equation}
%
Then, with $u = \cos\theta$ and $k_1 = \chi m_\mu$, one has
after the trivial integration over $\varphi$,
\begin{align}
\Gamma =& \; \frac{G_F^2 m_\mu^6}{12 \, \pi^3 \, (2 E_1)} \,
\int_0^1 \dd \xi \, \int_{-1}^1 \dd u \, 
\frac{ \xi^2 \, \sqrt{ 1 + \chi^2 } }%
{8 \, \sqrt{\xi \, \left( \xi (1 + u^2 \, \chi^2) +
2 u \sqrt{1 + \chi^2} \, \chi \right) + \chi^2 }}
\nonumber\\[0.1133ex]
& \; \times \left[ \left( 
4 \xi \, u \chi (\chi^2 + 1) + (4 \chi^2 + 3) \,
\sqrt{\chi^2 + 1} \right) \,
\sqrt{\xi \, \left( \xi (1 + u^2 \, \chi^2) +
2 u \sqrt{1 + \chi^2} \, \chi \right) + \chi^2 } \right.
\nonumber\\[0.1133ex]
& \; \left. - \xi \, u \, \sqrt{\chi^2 + 1} \, (8 \chi^2 + 7) \, \chi 
- 2 \xi^2 (\chi^2 + 1) \, (2 u^2 \chi^2 + 1) 
- 4 \chi^2 - 5 \chi^2 \right] \,.
\end{align}
After a somewhat tedious $u$ integration, the 
result can be written in terms of the variable
\begin{equation}
\label{cases}
\left| \xi \, \sqrt{1 + \chi^2} - \chi \right| =
\left\{ \begin{array}{cc}
\chi - \xi \, \sqrt{1 + \chi^2} & \qquad
0 < \xi < \dfrac{\chi}{\sqrt{1 + \chi^2} }
\nonumber\\[4ex]
\xi \, \sqrt{1 + \chi^2} - \chi & \qquad
\dfrac{\chi}{\sqrt{1 + \chi^2} } < \xi < 1
\end{array} \right. \,.
\end{equation}
The decay rate is naturally written as $\Gamma = \Gamma_1 + \Gamma_2$,
where the integration domains are such that 
$\left| \xi \, \sqrt{1 + \chi^2} - \chi \right|$ 
assumes either of the values given in Eq.~\eqref{cases}.
Here,
\begin{subequations} 
\begin{align} 
\Gamma_1 =& \;
\frac{G_F^2 m_\mu^6}{12 \, \pi^3 \, (2 E_1)} \,
\int\limits_0^{\chi/\sqrt{1 + \chi^2}} 
\frac{\dd \xi}{8 \chi}
 \xi \, \left( \sqrt{1 + \chi^2} \, \ln\left(\dfrac{1+\xi}{1-\xi}\right) 
- 2 \xi (1+\chi^2) \, 
\left( \chi \, \left(4 \chi \, (\sqrt{1+\chi^2}-\chi)-3 \right) +
\sqrt{1 + \chi^2} \right) \right)  \,,
\\[0.1133ex]
\Gamma_2 =& \;
\frac{G_F^2 m_\mu^6}{12 \, \pi^3 \, (2 E_1)} \,
\int\limits_{\chi/\sqrt{1 + \chi^2}}^1
\frac{\dd \xi}{8 \chi}
\xi \, \left( \sqrt{1 + \chi^2} \, 
\ln\left(\dfrac{1+\sqrt{1+\chi^2}}{1-\sqrt{1+\chi^2}}\right)
+ 2 \, \chi \, \left( 2 (1-\xi^2) \, \chi^2 - 2 \xi^2 + 3 \xi -1\right)
\, (1 + \chi^2) \right)  \,.
\end{align}
\end{subequations} 
The two contributions evaluate to the expressions,
\begin{subequations}
\begin{align}
\Gamma_1 = & \;
\frac{G_F^2 m_\mu^6}{12 \, \pi^3 \, (2 E_1)} \,
\frac{ 2 \chi (3 + 4 \chi^2) \, 
\left( \sqrt{1 + \chi^2} + 2 \chi^2 (\chi - \sqrt{1 + \chi^2}) - 
3 \, \ln\left( \dfrac{\sqrt{1 + \chi^2} + \chi}{\sqrt{1 + \chi^2} - \chi}
\right) \right)}{48 \, \chi \, \sqrt{1 + \chi^2}} \,,
\\[0.1133ex]
\Gamma_2 = & \;
\frac{G_F^2 m_\mu^6}{12 \, \pi^3 \, (2 E_1)} \,
\left[ \frac18 -
\frac{2 \chi (3 + 4 \chi^2) \, \left( \sqrt{1 + \chi^2} +
2 \chi^2 (\chi - \sqrt{1 + \chi^2}) -
3 \, \ln\left( \dfrac{\sqrt{1 + \chi^2} + \chi}{\sqrt{1 + \chi^2} - \chi}
\right) \right)}{48 \, \chi \, \sqrt{1 + \chi^2}} \right] \,,
\end{align}
\end{subequations}
so that 
\begin{equation}
\Gamma = \Gamma_1 + \Gamma_2 
= \frac{G_F^2 m_\mu^6}{12 \, \pi^3 \, (2 E_1)} \times \frac18 
= \frac{G_F^2 m_\mu^5}{192 \, \pi^3} \,\frac{m_\mu}{E_1} \,.
\end{equation}
This is the expected result for the muon decay width,
with the $1/E_1$ prefactor already included,
which is here obtained directly by an explicit
integration in the lab frame.

We have also verified~\cite{JeEh2016advhep} the results of Cohen and Glashow
for superluminal neutrino decay, based on the noncovariant
dispersion relation used in Ref.~\cite{CoGl2011},
via an independent calculation in the lab frame,
as envisaged in Ref.~\cite{BeLe2012}.
The treatment described in Ref.~\cite{CoGl2011}
is based on a Lorentz-violating dispersion relation
$E = |\vec k| \, v_\nu$ with $v_\nu > 1$, 
which constitutes a fundamentally different theoretical model
as compared to the Lorentz-invariant tachyonic
treatment presented here.
In addition to the decay rate,
we shall also consider the energy loss rate
of an incoming neutrino beam in the lab frame, due to LPCR and NPCR.
A remark is in order: 
The calculation of the energy loss per time of an incoming
muon beam, to complement a corresponding calculation
for the tachyonic neutrino beam, is not applicable,
because the end product of the decay is not a less energetic
muon, but the muon disappears from the beam altogether
(see Fig.~\ref{fig4}).

\begin{figure}[t!]
\begin{center}
\begin{minipage}{0.7\linewidth}
\begin{center}
\includegraphics[width=0.5\linewidth]{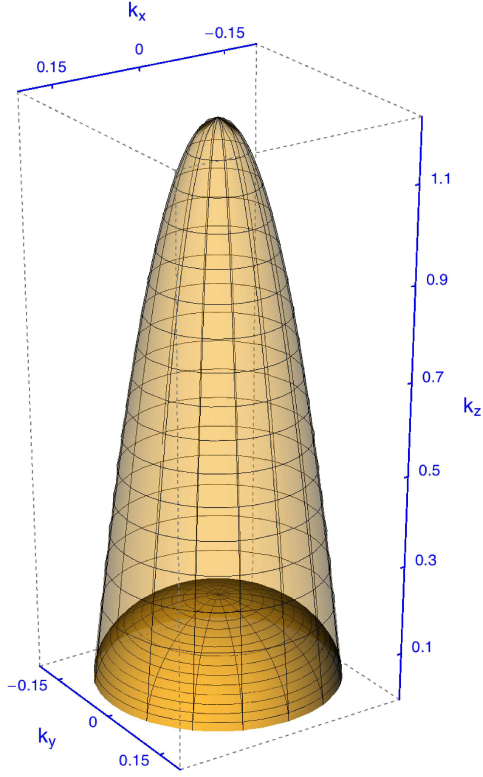}
\end{center}
\caption{\label{fig6} A tachyonic neutrino with 
wave vector $k_1 = 122$ and mass $-m^2_\nu = -(0.2)^2$ is incoming
along the positive $z$ direction. The
electron mass is set equal to unity, $m_e = 1$.
The threshold condition therefore reads as 
$q^2 = (p_1 - p_3)^2 \geq 4$. The  
boundaries of the volume of allowed $k_3$ vectors,
with Cartesian components
$k_x = k_{3x}$, $k_y = k_{3y}$, and $k_z = k_{3z}$,
are mainly concentrated in a narrow,
rotationally symmetric  cone about the $z$ axis.
Final states with $|\vec k_3| < m_\nu$ correspond
to evanescent outgoing waves, lead to a complex-valued
momentum transfer, and have to be excluded.}
\end{minipage}
\end{center}
\end{figure}

\begin{figure}[t!]
\begin{center}
\begin{minipage}{0.7\linewidth}
\begin{center}
\includegraphics[width=0.8\linewidth]{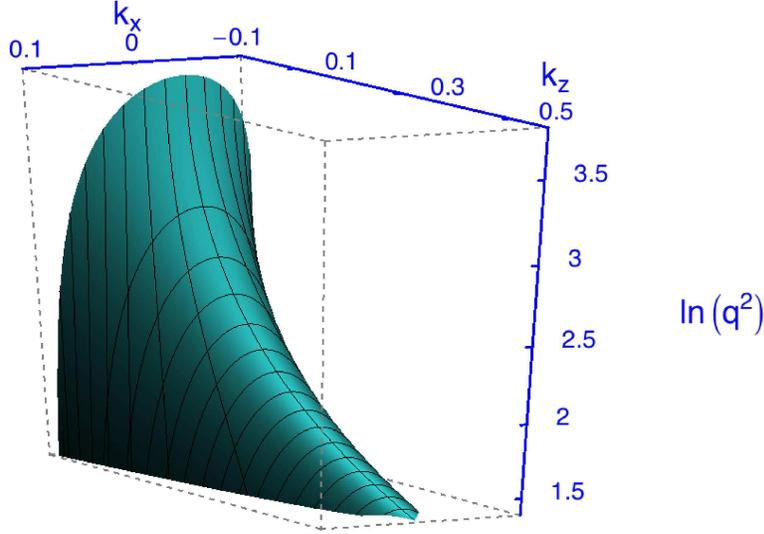}
\end{center}
\caption{\label{fig7} 
Plot of the tachyonic momentum
transfer $q^2$ in the kinematically
allowed region, for an oncoming neutrino
(along the positive $z$ direction) with parameters
$k_1 = 223$, $m_\nu = 0.1$, and $m_e = 1$.
The lines on the surface describe constant angle $\theta$.
and lines of constant $|\vec k_3|$.
We set $k_y=0$, which would otherwise correspond to 
an azimuth angle $\varphi = 0$ for the outgoing
neutrino momentum $\vec k_3$.
The vector modulus $k_3 = |\vec k_3|$ varies from its minimum tachyonic value of
$(k_3)_{\rm min} = m_\nu = 0.1$ to the maximum value
$(k_3)_{\rm max}$ given in Eq.~\eqref{k3max}.
For maximum $k_3$, $q^2$ assumes the threshold value $q^2 = 4 m_e^2 = 4$.
The maximum $q^2$ is reached at the minimum value for $k_3$, and for
collinear geometry, $\theta_3 = 0$, and reads as
$(q^2)_{\rm max} = 44.58$ [see Eq.~\eqref{q2max2}].}
\end{minipage}
\end{center}
\end{figure}

%
%
\subsection{Step 2: Tachyonic Neutrino Decay (Covariant Dispersion Relation)}
\label{step2}

We calculate the decay width of the incoming tachyonic neutrino,
in the lab frame, employing a relativistically covariant 
(tachyonic) dispersion relation,
with both incoming as well as outgoing neutrinos
on the tachyonic mass shell
[$E_1 = (\vec k_1^2 - m_\nu^2)^{1/2}$,
and $E_3 = (\vec k_1^2 - m_\nu^2)^{1/2}$, in the conventions
of Fig.~\ref{fig1}]. 
In the lab frame, in full accordance with Ref.~\cite{Jo2011},
the decay rate simply is
\begin{equation}
\label{GammaTacStart}
\Gamma = \frac{1}{2 E_1} \,
\int \frac{\dd^3 p_3}{(2 \pi)^3 \, 2 E_3}
\, \left(
\int \frac{\dd^3 p_2}{(2 \pi)^3 \, 2 E_2}
\int \frac{\dd^3 p_4}{(2 \pi)^3 \, 2 E_4}
(2 \pi)^4 \, \delta^{(4)}( p_1 - p_3 - p_2 - p_4 ) \,
\left[ {\widetilde \sum}_{\rm spins} | \calM |^2 \right] \right)
\end{equation}
Here, ${\widetilde \sum}_{\rm spins}$ refers to the 
specific way in which the average over the 
oncoming helicity states, and the outgoing helicities, 
needs to be carried out for tachyons.
As explained in Sec.~\ref{whywecannot},
we cannot reach the rest frame of the decaying particle
by a Lorentz transformation, in the case of a 
tachyonic neutrino.
Furthermore, as outlined in Sec.~\ref{maximum_momentum},
a calculation with just the Fermi effective coupling constant
actually is sufficient for the tachyonic case.
For the relevant interaction terms,
we use the same expression as in Sec.~\ref{step2}.
We recall the coupling of the 
decaying neutrino to the $Z^0$ boson according to Eq.~\eqref{L3},
\begin{equation}
\calL_3 = - \frac{g}{4 \cos\theta_W} \,
\bar \nu \, \gamma^\lambda \, (1 - \gamma^5) \, \nu \, Z_\lambda \,.
\end{equation}
For the coupling of the electron to the $Z$ boson,
we have according to Eq.~\eqref{L4},
\begin{equation}
\calL_4 = - \frac{g_w}{2 \, \cos\theta_W} \,
\overline e \, \left[ c_V \, \gamma^\lambda
- c_A \, \gamma^\lambda \, \gamma^5 \right] \, e \, Z_\lambda\,,
\qquad
c_V \approx 0 \,,
\qquad
c_A = -\frac12 \,.
\end{equation}
In view of the compensation mechanism
given by Eq.~\eqref{pref2},
the effective four-fermion Lagrangian thus is given by
\begin{equation}
\calL = \frac{G_F}{\sqrt{2}} \,
\left\{ \bar \nu \, \gamma^\lambda \, (1 - \gamma^5) \, \nu \right\} \,
\left\{ \overline e \, \left[ c_V \, \gamma^\rho
- c_A \, \gamma^\lambda\, \gamma^5 \right] \, e \right\} \,,
\end{equation}
The matrix element of the fundamental spinor solutions reads as 
follows,
\begin{equation}
\label{MM1}
\calM = \frac{G_F}{\sqrt{2}} \,
\left[ \overline u^\calT(p_3) \, \gamma_\lambda \, ( 1- \gamma^5) \,
u^\calT(p_1) \right] \,
\left[ \overline u(p_4) \, \left( c_V \gamma^\lambda -
c_A \, \gamma^\lambda \, \gamma^5 \right) \, v(p_2) \right] \,.
\end{equation}
Here, the $u^\calT(p_1)$, and $u^\calT(p_3)$, constitute Dirac 
spinor solutions of the tachyonic Dirac equation.
In the helicity basis~\cite{BeLiPi1982vol4,JeWu2013isrn},
denoted by a subscript $\sigma = \pm$,
the tachyonic particle and antiparticle spinors are
\begin{subequations}
\begin{align}
u^\calT_+(\vec k) = & \;
\left( \begin{array}{c} \sqrt{ | \vec k | + m} \; a_+(\vec k) \\
\sqrt{ | \vec k | - m} \; a_+(\vec k)
\end{array} \right)  \,,
\qquad
u^\calT_-(\vec k) =
\left( \begin{array}{c} \sqrt{ | \vec k | - m} \; a_-(\vec k) \\
- \sqrt{ | \vec k | + m} \; a_-(\vec k)
\end{array} \right)  \,,
\\[0.1133ex]
v^\calT_+(\vec k) = & \;
\left( \begin{array}{c} -\sqrt{ | \vec k | - m} \; a_+(\vec k) \\
-\sqrt{ | \vec k | + m} \; a_+(\vec k)
\end{array} \right)  \,,
\qquad
v^\calT_-(\vec k) =
\left( \begin{array}{c} - \sqrt{ | \vec k | + m} \; a_-(\vec k) \\
\sqrt{ | \vec k | - m} \; a_-(\vec k)
\end{array} \right)  \,,
\end{align}
\end{subequations}
where the $a_\sigma(\vec k)$ are the fundamental 
helicity spinors (see p.~87 of Ref.~\cite{ItZu1980}).

The properties of the tachyonic bispinor solutions
differ somewhat from those of the ``normal'' tardyonic 
bispinors. The well-known sum formula for the tardyonic 
states is ($\sigma$ denotes the spin orientation) 
\begin{equation}
\sum_\sigma u_\sigma(p) \otimes \overline u_\sigma(p) = 
\cancel p + m_e \,.
\end{equation}
For the tachyonic spin sums, one has the 
following sum rule for the positive-energy spinors,
\begin{equation}
\label{sumrule1}
\sum_\sigma (-\sigma )\;
u^\calT_\sigma(p) \otimes \overline u^\calT_\sigma(p) \, \gamma^5 =
\sum_\sigma \left(-\vec\Sigma \cdot \hat p \right)\;
u^\calT_\sigma(p) \otimes \overline u^\calT(p) \, \gamma^5 =
\cancel{p} - \gamma^5 \, m \,,
\end{equation}
where we use $p = (E, \vec p)$ as the 
convention for the four-momentum and 
$\hat p = \vec p / | \vec p|$ is the unit vector 
in the $\vec p$ direction.
Upon promotion to a four-vector, we have
$\hat p^\mu = (0, \hat p)$.
The sum rule can thus be reformulated as 
\begin{align}
\label{sumrule2}
\sum_\sigma u^\calT_\sigma(p) \otimes \overline u^\calT_\sigma(p)
=& \; \left( -\vec\Sigma \cdot \hat p \right) \,
( \cancel{p} - \gamma^5 \, m_\nu) \; \gamma^5
= - \gamma^5 \, \gamma^0 \, \gamma^i \, {\hat p}^i \;
( \cancel{p} - \gamma^5 \, m_\nu) \; \gamma^5
\nonumber\\[0.1133ex]
=& \; -\cancel{\tau} \, \gamma^5 \, \gamma^i \, \hat p_i \;
( \cancel{p} - \gamma^5 \, m_\nu) \; \gamma^5
= 
-\cancel{\tau} \, \gamma^5 \, \cancel{\hat{p}} \;
( \cancel{p} - \gamma^5 \, m_\nu) \; \gamma^5 \,,
\end{align}
where $\tau = (1,0,0,0)$ is a time-like unit vector.
In Refs.~\cite{JeWu2012epjc,JeWu2013isrn}, it has been 
established that a consistent formulation of the 
tachyonic propagator is achieved when we postulate that the 
right-handed neutrino states, and the left-handed 
antineutrino states, acquire a negative Fock-space norm
after quantization of the tachyonic spin-$1/2$ field.
Hence, in order to consider the decay process of 
an oncoming, left-handed, positive-energy neutrino, 
we should consider the projection onto negative-helicity 
states,
\begin{align}
\label{sumrule3}
\frac12 \, \left( 1 - \vec\Sigma \cdot \hat p \right) \;
\sum_\sigma u^\calT_\sigma(p) \otimes \overline u^\calT_\sigma(p) =
u_{\sigma=-1}(p) \otimes \overline u_{\sigma=-1}(p) =
\frac12 \, \left( 1 - \cancel{\tau} \, \gamma^5 \, \cancel{\hat{p}} \right) \,
( \cancel{p} - \gamma^5 \, m_\nu) \; \gamma^5 \,.
\end{align}
The squared and spin-summed matrix element is
\begin{align}
\label{tacTrace}
\mathop{{\widetilde \sum}}_{\rm spins} | \calM |^2 =& \;
\frac{G_F^2}{2} \,
{\rm Tr} \left[ 
\frac12 \, \left( 1 - \cancel{\tau} \, \gamma^5 \cancel{\hat{p}}_3 \right) \;
(\cancel{p}_3 - \gamma^5 \, m_{\nu}) \, \gamma^5 \, \gamma_\lambda \,
( 1- \gamma^5) 
\frac12 \, \left( 1 - \cancel{\tau} \, \gamma^5 \, \cancel{\hat{p}}_1 \right) \;
(\cancel{p}_1  - \gamma^5 \, m_{\nu} ) \, \gamma^5 \, \gamma_\nu  \,
( 1- \gamma^5) \right] \,
\nonumber\\[0.1133ex]
& \; \times {\rm Tr} \left[ ( \cancel{p}_4 + m_e ) \,
\left( c_V \gamma^\lambda - c_A \, \gamma^\lambda \, \gamma^5 \right) \,
( \cancel{p}_2 + m_e ) \,
\left( c_V \gamma^\rho - c_A \, \gamma^\rho \, \gamma^5 \right) \right]
= \frac{G_F^2}{2} \, \calS(p_1, p_2, p_3, p_4) \,,
\end{align}
with $c_V \approx 0$, $c_A \approx -1/2$ 
(the last step implicitly defines the expression $\calS$).
Here, the meaning of the notation~$\widetilde \sum_{\rm spins}$ becomes clear: 
We have summed over the spins of the outgoing electron-positron pair, 
while only one specific helicity is taken into account for 
the oncoming (decaying) neutrino.

The Dirac $\gamma$ traces in Eq.~\eqref{tacTrace} give rise 
to a rather lengthy expression, which can be simplified
somewhat because incoming and outgoing particles
are on their respective mass shells, $p_2^2 = p_4^2 = m_e^2$,
while on the tachyonic mass shell, we have
$p_1^2 = p_3^2 = -m_\nu^2$.
Some other scalar products vanish, e.g., 
the scalar product of the time-like unit
vector $\tau$ and the space-like unit vector, 
which is $\tau \cdot \hat p = 0$.

The result of the Dirac $\gamma$ traces from Eq.~\eqref{tacTrace}
can then be inserted into Eq.~\eqref{GammaTacStart},
and the $\dd^3 p_2$ and $\dd^3 p_4$ integrals 
can be carried out using the following formulas,
which we recall from Appendix~\ref{appb} 
[see Eq.~\eqref{resIJK}],
\begin{align}
I(q) =& \;
\int \frac{\dd^3 p_2}{2 E_2}
\int \frac{\dd^3 p_4}{ 2 E_4}
\delta^{(4)}( q - p_2 - p_4 ) =
\frac{\pi}{2} \, \sqrt{1 - \frac{4 \, m_e^2}{ q^2 }} \,,
\\[0.133ex]
J_{\lambda\rho}(q) =& \;
\int \frac{\dd^3 p_2}{2 E_2}
\int \frac{\dd^3 p_4}{ 2 E_4}
\delta^{(4)}( q - p_2 - p_4 ) \,
\left( p_{2 \lambda} \; p_{4 \rho} \right)
=
\sqrt{1 - \frac{4 \, m_e^2}{q^2}} \,
\left[ g_{\lambda \rho} \, \frac{\pi}{24} \,
\left( q^2 - 4 m_e^2 \right) +
q_\lambda \, q_\rho \,
\frac{\pi}{12} \,
\left( 1 + \frac{2 m_e^2}{q^2} \right) \right] \,,
\\[0.133ex]
K(q) =& \;
\int \frac{\dd^3 p_2}{2 E_2}
\int \frac{\dd^3 p_4}{ 2 E_4}
\delta^{(4)}( q - p_2 - p_4 ) \,
\left( p_2 \cdot p_4 \right)
= \frac{\pi}{4} \,
\sqrt{1 - \frac{4 \, m_e^2}{q^2}} \,
\left( q^2 - 2 m_e^2 \right) \,.
\end{align}
After the $\dd^3 p_2$ and $\dd^3 p_4$ integrations,
we are left with an expression of the form
\begin{align}
\label{GammaF}
\Gamma =& \; \frac{G_F^2}{2} \frac{1}{(2 \pi)^5} \,
\int\limits_{q^2 > 4 m_e^2} \frac{\dd^3 p_3}{2 E_3} \,
\calF(p_1, p_3) \,,
\end{align}
where 
\begin{align}
\label{defcalF}
\calF(p_1, p_3) =
\int \frac{\dd^3 p_2}{2 E_2}
\int \frac{\dd^3 p_4}{ 2 E_4}
\delta^{(4)}( p_1 - p_2 - p_3 - p_4 ) \, \calS(p_1, p_2, p_3, p_4) \,.
\end{align}
Both the expressions for $\calS(p_1, p_2, p_3, p_4)$ 
as well as $\calF(p_1, p_3)$ are too lengthy to be displayed
in the context of the current paper.  However, 
approximate formulas can be given, e.g., when the 
incoming energy $E_1$ is near threshold.

In order to obtain a better intuitive picture for the 
domain of allowed $p_3$ four-vectors, we have to 
analyze the tachyonic kinematics in some more detail. 
We calculate in the lab frame and assume that the 
oncoming neutrino has the energy-momentum four-vector
\begin{equation}
\label{p1}
p_1^\mu = (E_1, 0,0, k_1) \,,
\qquad
E_1 \geq (E_1)_{\thr} = 2 \frac{m_e}{m_\nu}  \, \sqrt{ m_e^2 + m_\nu^2} \,,
\qquad
k_1 \geq (k_1)_{\thr} = 2 \frac{m_e^2}{m_\nu} + m_\nu \,,
\qquad
E_1^2 - k_1^2 = -m_\nu^2
\end{equation}
[see Eqs.~\eqref{k1th} and~\eqref{E1th}].
The final-state four-vector is conveniently parameterized as
\begin{equation}
\label{p3}
p_3^\mu = (E_3, 
k_3 \, \sin\theta \, \cos\varphi , 
k_3 \, \sin\theta \, \sin\varphi , 
k_3 \, \cos\theta) \,,
\qquad
k_3 > m_\nu \,.
\qquad
E_3^2 - k_3^2 = -m_\nu^2
\end{equation}
The condition $k_3 > m_\nu$ is naturally imposed for 
tachyonic kinematics (see Fig.~\ref{fig3}).
The squared four-momentum transfer then reads as
\begin{align}
q^2 =& \; 2 \, \left( 
\sqrt{E_1^2 + m_\nu^2} \sqrt{E_3^2 + m_\nu^2} \, \cos \theta 
-E_1 E_3 - m_\nu^2 \right) 
\nonumber\\[0.1133ex]
=& \; 2 \, \left(
k_1 \, k_3 \, \cos \theta
-\sqrt{k_1^2 - m_\nu^2} \, \sqrt{k_3^2 - m_\nu^2} - m_\nu^2 \right) \,,
\end{align}
where it is convenient to define $u = \cos\theta$.
One may solve for the threshold angle $\cos\theta_{\thr}$,
\begin{equation}
q^2 = 4 m_e^2 
\quad
\Rightarrow
\quad
u = \cos\theta = u_{\thr} = \cos\theta_{\thr} = 
\frac{E_1 \, E_3 + 2 m_e^2 + m_\nu^2}%
{ \sqrt{E_1^2 + m_\nu^2} \sqrt{E_2^2 + m_\nu^2} } \,.
\end{equation}
For given $E_1$ and $E_3$, all angles $\theta$ with
$\cos\theta > \cos\theta_{\thr}$, i.e.,
for $\theta < \theta_{\thr}$, are permissible.
Conversely, setting $\cos\theta_{\thr} = 1$ and $E_3 = 0$,
one may solve for $E_1$ and rederive the threshold 
condition~\eqref{k1th}.
All of this implies that the domain of permissible
$\vec k_3$ vectors, near threshold, is centered 
about the $z$ axis and forms a ``cupola'' of 
inner radius $m_\nu$ (see Fig.~\ref{fig6}).
Within the kinematically allowed region,
the tachyonic momentum transfer $q^2$ is 
plotted in Fig.~\ref{fig7}.

For given $E_1$, the widest opening angle $\theta = \theta_{\thr}$
is reached when $E_3$ becomes zero. One finds
\begin{equation}
\cos\theta_{\thr} \bigg|_{E_3 = 0} =
\frac{2 m_e^2 + m_\nu^2}{m_\nu \, \sqrt{E_1^2 + m_\nu^2}}
\approx 
\left( \frac{2 m_e^2}{m_\nu} + m_\nu \right) \, E_1^{-1}
= \frac{k_{\thr}}{E_1} \,,
\end{equation}
where the latter form is valid in the high-energy limit.
Here, $k_{\thr} = \frac{2 m_e^2}{m_\nu} + m_\nu$ is the
threshold momentum.
It means that the produced pairs will be emitted in a
very narrow cone, centered in the forward direction
with respect to the decaying neutrino.

Maximum squared momentum transfer is reached at 
$E_3 = 0$ and $\theta = 0$,
\begin{equation} 
\label{q2max2}
q^2 = q^2_{\rm max} 
= 2 \, m_\nu \, \left( \sqrt{E_1^2 + m_\nu^2} - m_\nu \right)  
= 2 \, m_\nu \, \left( k_1 - m_\nu \right)  \,,
\end{equation}
confirming Eq.~\eqref{q2max}. Maximum outgoing energy 
$E_3$ is reached for minimum momentum transfer $q^2 = q^2_{\rm min} = 
4 m_e^2$, with the final-state neutrino propagating into the 
positive $z$ direction. Its energy is
\begin{equation}
\label{E3max}
(E_3)_{\rm max} = 
E_1 \, \left( 2 \, \frac{m_e^2}{m_\nu^2} + 1 \right) 
- \frac{2 m_e \, \sqrt{E_1^2 + m_\nu^2} \, \sqrt{m_e^2 + m_\nu^2}}{m_\nu^2}
\approx \frac{E_1 \, m_\nu^2}{4 m_e^2} \,,
\end{equation}
where the latter form is valid in the limit of large $E_1$.
This corresponds to the maximum allowed $k_3$,
\begin{equation}
\label{k3max}
(k_3)_{\rm max} =
\frac{k_1 (2 m_e^2+m_\nu^2) -2 m_e \sqrt{k_1^2-m_\nu^2} \, \sqrt{m_e^2 + m_\nu^2}}%
{m_\nu^2}
\approx \frac{k_1 \, m_\nu^2}{4 m_e^2} \,,
\end{equation}
where, again, the latter form is valid in high-energy limit $k_1 \to \infty$.
A plot of the physically relevant range for $q^2$ is given in Fig.~\ref{fig7}.

Because of azimuthal symmetry, it is easily possible to find
a convenient parameterization of the regime of allowed $\vec k_3$
in spherical coordinates [using the parameterization given in Eq.~\eqref{p3}].
We may finally express the integrand $\calF$ from Eq.~\eqref{defcalF}
in terms of the initial and final energies $E_1$ and $E_3$ of the 
decay process, and of the scattering angle $\theta$
(with $u = \cos\theta$).
The decay rate given by Eq.~\eqref{GammaF} can thus be 
written as follows,
\begin{align}
\label{cfGamma}
\Gamma =& \; \frac{G_F^2}{2} \frac{1}{(2 \pi)^5} \,
\int\limits_0^{2 \pi} \dd \varphi \,
\int\limits_{(k_3)_{\rm min}}^{(k_3)_{\rm max}} 
\frac{\dd k_3 \, k_3^2}{2 E_3} \,
\int\limits_{u_{\thr}}^1 \dd u \, \calF(E_1, E_3, u) 
\nonumber\\[0.1133ex]
=& \; \frac{G_F^2}{4} \frac{1}{(2 \pi)^4} \,
\int\limits_{(E_3)_{\rm min}}^{(E_3)_{\rm max}}
\dd E_3 \, \sqrt{E_3^2 + m_\nu^2}
\int\limits_{u_{\thr}}^1 \dd u \, \calF(E_1, E_3, u) \,.
\end{align}
Here, $(k_3)_{\rm min} = m_\nu$, while $(k_3)_{\rm max}$ is 
given by Eq.~\eqref{k3max}.
Furthermore, we have $(E_3)_{\rm min} = 0$, while $(E_3)_{\rm max}$ is 
given by Eq.~\eqref{E3max},
and we have used the identity
\begin{equation}
\dd k_3 \, k_3 = \dd E_3 \, E_3 \,,
\qquad
k_3 = \sqrt{E_3^2 + m_\nu^2} \,.
\end{equation}
It is instructive to consider the 
double-differential energy loss $\dd^2 E_1$, 
for a particle traveling at velocity $v_\nu \approx c$
(we restore factors of $c$ for the moment),
as it undergoes a decay with energy loss $E_1 - E_3$,
due to the energy-resolved decay rate
$(\dd \Gamma/\dd E ) \, \dd E$, in time $\dd t = \dd x/c$.
It reads as follows,
\begin{equation}
\dd^2 E_1 = -(E_1 - E_3) \, \frac{\dd \Gamma}{\dd E_3}
\dd E_3 \, \frac{\dd x}{c} \,.
\end{equation}
Now we revert to natural units with $c = 1$, 
divide both sides of the equation
by $\dd x$ and integrate over the energy loss.
One obtains
\begin{equation}
\frac{\dd E_1}{\dd x} =
- \int \dd E_3 \, (E_1 - E_3) \, \frac{\dd \Gamma}{\dd E_3} \,.
\end{equation}
Hence, the energy loss rate is obtained as
\begin{equation}
\label{cfdEdx}
\frac{\dd E}{\dd x} = -\frac{G_F^2}{4} \frac{1}{(2 \pi)^4} \,
\int\limits_{(E_3)_{\rm min}}^{(E_3)_{\rm max}}
\dd E_3 \sqrt{E_3^2 + m_\nu^2} \, (E_1 - E_3) \,
\int\limits_{u_{\thr}}^1 \dd u \, \calF(E_1, E_3, u) \,.
\end{equation}
After a long, and somewhat tedious integration one finds the 
following expressions,
which have been briefly indicated in Ref.~\cite{JeEh2016advhep},
\begin{subequations}
\label{GammadEdx1}
\begin{align}
\Gamma =& \; \left\{ \begin{array}{cc}
\dfrac32 \; \dfrac{G_F^2 \, m_\nu^5}{192 \, \pi^3} \;
\dfrac{m_\nu \, (E_1 - E_{\thr})^2}{m_e^2 \, E_{\thr}} & 
\qquad E_1 \gtrapprox E_{\thr} \\[4ex]
\dfrac23 \; \dfrac{G_F^2 \, m_\nu^5}{192 \pi^3} \; \dfrac{E_1}{m_\nu} &
\qquad E_1 \gg E_{\thr} 
\end{array}
\right. \,,
\\[4ex]
\frac{\dd E}{\dd x} =& \; \left\{ \begin{array}{cc}
3 \, \dfrac{G_F^2 \, m_\nu^5}{192 \, \pi^3 } \;
\dfrac{(E_1 - E_{\thr})^2}{E_{\thr}} &
\qquad E_1 \gtrapprox E_{\thr} \\[4ex]
\dfrac43 \; \dfrac{G_F^2 \, m_\nu^5}{192 \pi^3} \, \dfrac{E^2_1}{m_\nu} &
\qquad E_1 \gg E_{\thr}
\end{array}
\right. \,.
\end{align}
\end{subequations}
In the high-energy limit, one may rewrite the expressions
as follows,
\begin{equation}
\label{GammadEdx2}
\Gamma =
\frac{G_F^2 \, E_1^5 \, \delta^2}{288 \, \pi^3} \,,
\qquad
\qquad
\frac{\dd E}{\dd x} =
\frac{G_F^2 \, E_1^6 \, \delta^2}{144 \, \pi^3} \,,
\qquad
\qquad
E_1 \gg E_{\thr} \,.
\end{equation}
The ratio of the energy loss rate to the decay rate is 
given as
\begin{equation}
\label{GammadEdx3}
\frac{1}{\Gamma} \, 
\frac{\dd E}{\dd x} = \left\{ \begin{array}{cc}
2 \, \dfrac{m_e^2}{m_\nu} \approx E_{\thr} & \qquad E_1 \geq E_{\thr} \\[4ex]
2 E_1 & \qquad E_1 \gg E_{\thr}
\end{array}
\right. \,.
\end{equation}
Here, according to Eq.~\eqref{E1th}, the threshold energy is 
\begin{equation}
E_{\thr} = (E_1)_{\thr} =
\sqrt{ (k_1)^2_{\thr} - m_\nu^2 } = 
2 \, \frac{m_e}{m_\nu}  \, \sqrt{ m_e^2 + m_\nu^2} 
\approx 
2 \, \dfrac{m_e^2}{m_\nu} \,.
\end{equation}
For all results given in Eqs.~\eqref{GammadEdx1},~\eqref{GammadEdx2}
and~\eqref{GammadEdx3}, we have assumed that $m_e \gg m_\nu$.
Interpolating formulas are given as
\begin{align}
\Gamma \approx & \;
\dfrac{G_F^2 \, m_\nu^6}{128 \, \pi^3 \, m_e^2}
\dfrac{(E_1 - E_{\thr})^2}{E_{\thr}} \,
\left( 1+ \frac{9 m_\nu^2}{4 m_e^2} \frac{E_1 - E_{\thr}}{E_{\thr}} \right)^{-1}
\,,
\nonumber\\[0.1133ex]
\frac{\dd E}{\dd x} \approx & \;
\dfrac{G_F^2 m_\nu^5}{64 \pi^3 }
\dfrac{(E_1 - E_{{\thr}})^2}{E_{{\thr}}} \,
\left( \frac{4 E_1 E_{\thr}}{4 E_{\thr}^2 + 9 m_\nu (E_1 - E_{\thr})} \right) \,.
\end{align}
These formulas interpolate between the regimes
$E_1 \gtrapprox E_{\thr}$ and 
$E_1 \gg E_{\thr}$ given in Eq.~\eqref{GammadEdx1}.

%
%
\section{Neutrino Pair Cerenkov Radiation}
\label{sec4}

%
%
\subsection{Preliminary Steps}

Having laid out the formalism in Sec.~\ref{sec3},
we can be brief in the current section.
In the lab frame, again, the decay rate evaluates to
\begin{align}
\label{GammaTacStartNPCR}
\Gamma =& \;  \frac{1}{2 E_1} \,
\int \frac{\dd^3 p_3}{(2 \pi)^3 \, 2 E_3}
\, \left(
\int \frac{\dd^3 p_2}{(2 \pi)^3 \, 2 E_2}
\int \frac{\dd^3 p_4}{(2 \pi)^3 \, 2 E_4}
\right.
\nonumber\\
& \; \left. \times (2 \pi)^4 \, \delta^{(4)}( p_1 - p_3 - p_2 - p_4 ) \,
\left[ {\widetilde \sum}_{\rm spins} | \calM |^2 \right] 
\right) \,.
\end{align}
Here, ${\widetilde \sum}_{\rm spins}$ refers to the
specific way in which the average over the
oncoming helicity states, and the outgoing helicities,
needs to be carried out for tachyons~\cite{JeEh2016advhep}.
We use the Lagrangian~\eqref{L3}.
The effective four-fermion interaction thus is
\begin{equation}
\calL = \frac{G_F}{2 \sqrt{2}} \,
\left[ \overline \nu \, 
\gamma^\mu (1 - \gamma^5) \, \nu \right] \,
\left[ \overline \nu \, 
\gamma^\mu (1 - \gamma^5) \, \nu \right] \,.
\end{equation}
The matrix element $\calM$ evaluates to
\begin{equation}
\label{MM2}
\calM = \frac{G_F}{2 \sqrt{2}} \,
\left[ \overline u^\calT(p_3) \, \gamma_\lambda \, ( 1- \gamma^5) \,
u^\calT(p_1) \right] \,
\left[ \overline u^\calT(p_4) \, \gamma^\lambda \, 
( 1 - \gamma^5 ) \, v^\calT(p_2) \right] \,,
\end{equation}
in the notation for the tachyonic bispinors adopted
previously. We use, again, the helicity-projected sum rule
\begin{equation}
\label{sr2}
\sum_\sigma u^\calT_\sigma(p) \otimes \overline u^\calT_\sigma(p)
= \left( -\vec\Sigma \cdot \hat k \right) \,
( \cancel{p} - \gamma^5 \, m_\nu) \; \gamma^5 
= - \gamma^5 \, \gamma^0 \, \gamma^i \, {\hat k}^i \;
( \cancel{p} - \gamma^5 \, m_\nu) \; \gamma^5 \\
= -\cancel{\tau} \, \gamma^5 \, \cancel{\hat{k}} \;
( \cancel{p} - \gamma^5 \, m_\nu) \; \gamma^5 \,,
\end{equation}
where $\tau = (1,0,0,0)$ is a time-like unit vector.
This leads to 
\begin{equation}
\label{sr3}
\frac12 \, \left( 1 - \vec\Sigma \cdot \hat k \right) \;
\sum_\sigma u^\calT_\sigma(\vec k) \otimes 
\overline u^\calT_\sigma(\vec k) 
= u_{\sigma=-1}(p) \otimes \overline u_{\sigma=-1}(p) 
= \frac12 \, \left( 1 - \cancel{\tau} \, \gamma^5 \, 
\cancel{\hat{k}} \right) \,
( \cancel{p} - \gamma^5 \, m_\nu) \; \gamma^5 \,.
\end{equation}
The squared and spin-summed matrix element 
for the tachyonic decay process thus is
\begin{align}
\label{tacTraceNPCR}
\mathop{{\widetilde \sum}}_{\rm spins} | \calM |^2 =& \;
\frac{G_F^2}{8} \, {\rm Tr} \left[ 
\frac12 \, \left( 1 - \cancel{\tau} \, \gamma^5 \cancel{\hat{k}}_3 \right) \;
(\cancel{p}_3 - \gamma^5 \, m_{\nu}) \, \gamma^5 \, \gamma_\lambda \,
( 1- \gamma^5) \, \frac12 \, 
\left( 1 - \cancel{\tau} \, \gamma^5 \, \cancel{\hat{k}}_1 \right) \;
(\cancel{p}_1  - \gamma^5 \, m_{\nu} ) \, \gamma^5 \, \gamma_\nu  \,
( 1- \gamma^5) \right] \,
\nonumber\\[0.1133ex]
& \; \times {\rm Tr} \left[ 
\frac12 \, 
\left( 1 - \cancel{\tau} \, \gamma^5 \cancel{\hat{k}}_4 \right) \,
( \cancel{p}_4 - \gamma^5 \, m_\nu ) \, \gamma^5 \,
\gamma^\lambda ( 1 - \gamma^5 ) 
\times \frac12 \, 
\left( 1 - \cancel{\tau} \, \gamma^5 \cancel{\hat{k}}_2 \right) \,
( \cancel{p}_2 + \gamma^5 \, m_\nu ) \, \gamma^5 \,
\gamma_\lambda ( 1 - \gamma^5 ) 
\right] \,.
\end{align}
Again, we have chosen the convention to denote the 
by $p_2$ the momentum of the outgoing antiparticle.

%
%
\subsection{Integration and Results}

We now turn to the integration over the 
four-momenta of the outgoing particles.
In the calculation, one may use the fact 
that the helicity projector is well 
approximated equal to the 
chirality projector for tachyonic particle 
in the high-energy limit (with the energy 
being significantly larger than the tachyonic mass).
On the tachyonic mass shell, one has
$p_1^2 = p_2^2 = p_3^2 = p_4^2 = -m_\nu^2$.
The trace over the Dirac $\gamma$ matrices
can be evaluated with standard computer algebra~\cite{HsYe1992,Wo1999}
and is inserted into Eq.~\eqref{GammaTacStart}.
The $\dd^3 p_2$ and $\dd^3 p_4$ integrals are carried 
out with the help of the formulas~\eqref{resI},~\eqref{resJ} 
and~\eqref{resK}, under the appropriate replacement 
$m_e^2 \to -m_\nu^2$.
After the $\dd^3 p_2$ and $\dd^3 p_4$ integrations,
we are left with an expression of the form
\begin{align}
\label{GammaF_NPCR}
\Gamma =& \; \frac{G_F^2}{8} \frac{1}{(2 \pi)^5} \,
\int\limits_{q^2 > 4 m_e^2} \frac{\dd^3 p_3}{2 E_3} \,
{\overline \calF}(p_1, p_3) \,,
\end{align}
where
\begin{equation}
\label{defoverlinecalF}
{\overline \calF}(p_1, p_3) =
\int \frac{\dd^3 p_2}{2 E_2}
\int \frac{\dd^3 p_4}{ 2 E_4}
\delta^{(4)}( p_1 - p_2 - p_3 - p_4 ) \, 
{\overline \calS}(p_1, p_2, p_3, p_4) \,.
\end{equation}
The expressions for ${\overline \calS}(p_1, p_2, p_3, p_4)$
as well as ${\overline \calF}(p_1, p_3)$ are too lengthy to be displayed
in the context of the current paper. 
We assume the same kinematics as in Eqs.~\eqref{p1} and~\eqref{p3}.
The integrations are done
with under the conditions that all $0 < E_3 < E_1$, 
and all $q^2 = (p_2 + p_4)^2$ for the pair are allowed
[see Eq.~\ref{q2range}], leading to 
\begin{align}
\Gamma =& \; \frac{G_F^2}{8} \frac{1}{(2 \pi)^5} \,
\int\limits_0^{2 \pi} \dd \varphi \,
\int\limits_{k_3 = m_\nu}^{k_{\rm max}}
\frac{\dd k_3 \, k_3^2}{2 E_3} \,
\int\limits_{-1}^1 \dd u \, {\overline \calF}(E_1, E_3, u) 
\nonumber\\[0.1133ex]
=& \; \frac{G_F^2}{16} \frac{1}{(2 \pi)^4} \,
\int\limits_{0}^{E_1}
\dd E_3 \, \sqrt{E_3^2 + m_\nu^2}
\int\limits_{-1}^1 \dd u \, {\overline \calF}(E_1, E_3, u) \,,
\end{align}
in full analogy with Eq.~\eqref{q2range}.
Finally, the energy loss rate is obtained 
in full analogy with Eq.~\eqref{cfdEdx},
\begin{equation}
\frac{\dd E}{\dd x} =
 -\frac{G_F^2}{4} \frac{1}{(2 \pi)^4} \,
\int\limits_{0}^{E_1}
\dd E_3 \sqrt{E_3^2 + m_\nu^2} \, (E_1 - E_3) \,
\int\limits_{-1}^1 \dd u \, \calF(E_1, E_3, u) \,.
\end{equation}
After rather tedious integration one finds the
following expressions,
\begin{subequations}
\label{GammadEdxNPCR}
\begin{align}
\label{GammaNPCR}
\Gamma =& \; 
\dfrac13 \; \dfrac{G_F^2 \, m_\nu^4}{192 \pi^3} \; E_1 \,,
\\[4ex]
\label{dEdxNPCR}
\frac{\dd E_1}{\dd x} =& \; 
\dfrac13 \; \dfrac{G_F^2 \, m_\nu^4}{192 \pi^3} \; E^2_1 \,.
\end{align}
\end{subequations}
Strictly speaking, these formulas are valid 
only for $E_1 \gg m_\nu$, but this condition is 
easily fulfilled for all phenomenologically 
relevant neutrino energy, assuming that 
the modulus of neutrino masses $| m_\nu |$ 
does not exceed $1 \, {\rm eV}$.
We reemphasize that, 
unlike in Eq.~\eqref{GammadEdx1}, 
there is no further threshold condition.
Parametrically, the results in 
Eqs.~\eqref{GammadEdx1} and~\eqref{GammadEdxNPCR}
are of the same order-of-magnitude.
Hence, neutrino pair emission is the dominant decay channel
in the medium-energy domain, for an oncoming tachyonic neutrino 
flavor eigenstate.

%
%
\section{Phenomenological Consequences}
\label{sec5}

%
%
\subsection{Decay Processes on Cosmic Distance and Time Scales}

If we assume that the tachyonic neutrino hypothesis is real,
then a natural question to ask concerns the phenomenological consequences 
of the calculations outlined above.
Parametrically, the decays by LPCR and NPCR 
described by Eqs.~\eqref{GammadEdx1} and~\eqref{GammadEdxNPCR}
might set important limits on the observability of 
tachyonic neutrinos, provided the absolute 
magnitude of the decay energy loss rates
are sufficiently large in order to induce a 
significant decay probability for 
neutrinos traveling across the Universe.
This is because neutrinos registered by IceCube have to
``survive'' the possibility of energy loss by decay,
and if they are tachyonic, then lepton and neutrino
pair Cerenkov radiation processes become kinematically
allowed.

Indeed, it is known that even very small 
Lorentz-violating parameters in a Lorentz-violating
extension of the standard model 
may induce very significant energy loss processes
at high energies~\cite{StSc2014,St2014}.
This is because at high energies, 
small violations of the Lorentz symmetry correspond 
to very high virtualities of the particles
(kinematic deviations from the mass shell),
and therefore, the magnitude of the 
Lorentz-violating parameters is in fact severely 
constrained by the 37 neutrinos with $E> \SI{60}{\TeV}$
which are believed to be of cosmological origin
and which have been registered by the IceCube
collaboration~\cite{AaEtAl2013,AaEtAl2014}.
Meanwhile, preliminary evidence for a
through-going muon depositing an energy of $\geq (2.6 \pm 0.3) \, {\rm PeV}$
has been presented by some members of the IceCube collaboration~\cite{EV1}.
The event could be interpreted in terms of a decay product of a neutrino of even higher
energy~\cite{EV1}. (The difference of the energy deposited inside
the detector and the neutrino energy, according to
Fig.~4 of Ref.~\cite{AaEtAl2014}, is small.)
If confirmed, this event would lead to 
even more restrictive bounds on the Lorentz-violating 
parameters.

The results given in Eqs.~\eqref{GammadEdxNPCR} 
for the decay rate and energy loss rate due to 
NPCR are not subject to a threshold energy.
Parametrically, they are of the same order-of-magnitude 
as those given for lepton pair Cerenkov radiation in 
Eq.~\eqref{GammadEdx1}, but the threshold energy is zero.
Let us take as a typical cosmological distance
15~billion light years,
\begin{equation}
L = 15 \times 10^9 \, {\rm ly} = 1.42 \times 10^{26} \, {\rm m}  \,,
\end{equation}
and assume a (relative large) neutrino mass
parameter of $m_0 = 10^{-2} \, {\rm eV}$.
One obtains for the relative energy loss
according to Eq.~\eqref{GammaNPCR},
\begin{equation}
\frac{L}{E_1} \, \frac{\dd E_1}{\dd x} = 
\dfrac13 \; \dfrac{G_F^2 \, m_0^4}{192 \pi^3} \, E_1 \, L =
5.02 \times 10^{-20} \, \frac{E_1}{{\rm MeV}} \,.
\end{equation}
Even at the large ``Big Bird'' 
energy of $E_\nu \approx 2\,{\rm PeV}$,
the relative energy loss over 15 billion 
light years does not exceed $5 \times 10^{-20}$.

Again, assuming that $m_\nu = 10^{-2} \, {\rm eV}$),
one obtains for the decay rate the result
\begin{equation}
\Gamma =
\dfrac13 \; \dfrac{G_F^2 \, m_0^4}{192 \pi^3} \; E_1 
= 1.06 \times 10^{-37} \, \left( \frac{E_1}{{\rm MeV}} \right) \, 
\left( \frac{\rm rad}{\rm s} \right) \,.
\end{equation}
Even for $E_\nu \approx 2\,{\rm PeV}$, this means that the 
decay rate is only of order $10^{-28} \, \frac{\rm rad}{\rm s}$,
corresponding to a lifetime of 
$\sim 10^{20}$~years, far exceeding the age
of the Universe. Within the tachyonic model,
quite surprisingly, both LPCR as well as NPCR
are phenomenologically irrelevant, even for the 
highest-energy neutrinos registered by IceCube.

%
%
\subsection{Neutrino Mass and Flavor Eigenstates}

The tachyonic Dirac equation~\cite{ChHaKo1985,JeWu2012epjc} reads as
\begin{equation}
\label{tacdirac}
( \ii \gamma^\mu \partial_\mu - \gamma^5 m_\nu ) \psi(x) = 0 \,,
\qquad
( \ii \gamma^\mu p_\mu - \gamma^5 m_\nu ) u^\calT(p) = 0 \,,
\end{equation}
where the latter form holds for the plane-wave {\em ansatz}
$\psi(x) =  u^\calT(p) \, \exp(-\ii p \cdot x)$.
The bispinor solutions $u^\calT(p)$ have 
been discussed at length in Refs.~\cite{JeWu2012epjc,JeWu2013isrn}
and are used here in Eqs.~\eqref{MM1} and~\eqref{MM2}.
They apply, first and foremost, to a mass eigenstate,
with a definite tachyonic mass parameter.
The Fermi couplings are universal among all neutrino
flavors, and hence, the interaction Lagrangians used in our 
paper share this property. 
Our results~\eqref{GammadEdx1} and~\eqref{GammadEdxNPCR}
for the decay and energy loss rates thus apply, at face value, to
an incoming neutrino mass eigenstate. The results are thus
relevant to the non-sterile neutrino flavor
if at least one of the three observed non-sterile 
neutrino mass eigenstates is tachyonic. 

We recall that the flavor eigenstates 
$\nu_f$ are connected to the mass eigenstates $m_i$ by the 
Pontecorvo--Maki--Nakagawa--Sakata (PMNS) matrix,
\begin{equation}
m^2(\nu_f) = \sum_{i=1}^3 |U_{fi}|^2 \, m_i^2 \,,
\end{equation}
where $U_{fi}$ denote the elements of the 
flavor-mass mixing matrix.
The decay and energy loss rates of the flavor
eigenstates are given as
\begin{align}
\Gamma(\nu_f) =& \; \sum_{i=1}^3 |U_{fi}|^2 \, \Gamma(m_i) \,,
\\[0.1133ex]
\frac{\dd E}{\dd x}(\nu_f) =& \;
\sum_{i=1}^3 |U_{fi}|^2 \;\; \frac{\dd E}{\dd x}(m_i) \,.
\end{align}
For a slower-than light mass eigenstate $i$,
one sets $\Gamma(m_i)$ and 
$\frac{\dd E}{\dd x}(m_i)$ to zero.
Here, just to be pedantic, we should point out that the 
calculation of NPCR in this case has to be modified 
to include all tachyonic mass eigenstates in the 
exit channel, conceivably modifying the overall
results as much as by adding a multiplicative factor three
(if all mass eigenstates are available in the exit channel
of the tachyonic NPCR decay, see Fig.~\ref{fig1}(b)). 

%
%
\subsection{Lorentz Invariance}
\label{LIV}

The tachyonic dispersion relation $E^2 - \vec k^{\,2} = -m_\nu^2$
conserves Lorentz invariance.  Hence, one might ask about the
Lorentz invariance of our results, 
and in particular, about the Lorentz invariance 
of the threshold condition~\eqref{Eth_form}; finally, one might
``chase'' the high-energy neutrino, lowering its energy in the
Lorentz-transformed, moving frame to a value below threshold.  This question
finds an answer in the subtleties of the tachyonic theory; we 
follow the discussion in Ref.~\cite{Fe1967}.
Namely, upon a Lorentz transformation of the vacuum state,
because there is no ``energy mass gap'' between the positive- and
negative-energy states, some of the annihilation operators of quantized fields
will turn into creation operators, and vice versa.  This point is explained in
detail around Eqs. (4.7)---(4.9) of Ref.~\cite{Fe1967}.  (Incidentally, it is
observed at the same place that the fundamental creation and annihilation
operators of tachyonic fields have to be quantized according to fermionic
statistics, which is another argument in favor of spin-$1/2$ rather than
spinless tachyonic theories.) Furthermore, around Eq.~(5.7) of Ref.~\cite{Fe1967},
it is argued that the vacuum state in a tachyonic theory cannot be Lorentz
invariant, but is filled with those (real) anti-fermions whose energies are
``pushed down'' to energies below zero, 
from initially positive-energy states, under the
Lorentz transformation.  Our Fig.~\ref{fig2} illustrates how the decay and energy loss
rates, under a Lorentz transformation, turn into neutrino-antineutrino
collision rates (leading to decay and energy loss) with the ``downshifted''
real antiparticle states which are the result of the Lorentz transformation,
finally restoring the Lorentz invariance of the 
results for the decay and energy loss rates, given in 
Eqs.~\eqref{GammadEdx1} and~\eqref{GammadEdxNPCR}.

%
%
\subsection{Superluminal Signal Propagation}

A very important question regarding the conceivable
existence of tachyonic neutrinos concerns the 
possibility of superluminal signal propagation.
We thus follow Appendix A of Ref.~\cite{JeEtAl2014} and ask 
how difficult it is to 
reliably ``stamp'' any information onto the superluminal 
neutrinos. When assuming the dispersion relation 
$E = (\vec k^{\,2} - m_\nu^2)^{1/2}$
with its classical equivalent $E = m_\nu/\sqrt{v_\nu^2 - 1} = 0$,
the dilemma is that 
high-energy tachyonic neutrinos approach the light cone
and travel only infinitesimally faster than light itself.
In the high-energy limit,
their interaction cross sections may be sufficiently large to 
allow for good detection efficiency but this is achieved at the 
cost of sacrificing the ``speed advantage'' in comparison
to the speed of light. Low-energy tachyonic 
neutrinos may a substantially faster than light,
but their interaction cross sections are small and the information 
sent via them may be  lost. 
The smallness of the cross sections sets important boundaries
for the possibility to transmit information, as follows.
In Appendix~A of Ref.~\cite{JeEtAl2014}, it has been
shown that, by postulating that 
superluminal particles should not have the capacity 
to transport any ``imprinted'' information into the past,
one is naturally led to the assumption that any 
conceivable superluminal particles have to be
very light, and weakly interacting.

Following Fig.~1 of Ref.~\cite{FoZe2012} and Ref.~\cite{OULU_WEB2},
we now supplement these considerations with a numerical estimate.
Neutrino-electron cross sections for 
$1 \, {\rm GeV} < E_\nu < 1 \, {\rm PeV}$ can be 
estimated to good accuracy using the formula
\begin{equation}
\label{oom}
\sigma = A_0 \, \frac{E_\nu}{E_0} \,,
\end{equation}
with $A_0 \sim 0.0095$\,fb and $E_0 = 1$\,GeV.
By order-of-magnitude, Eq.~\eqref{oom} remains valid for
neutrino scattering off electrons, for all three neutrino flavors, even
if additional charged-current interactions exist for electron
neutrinos, due to exchange graphs with virtual $W$ bosons (for muon and
tau neutrinos, only the $Z$ boson contributes at tree level). 
A particle
typically cannot be localized to better than an area equal to the
square of its (reduced) Compton wavelength
(we temporarily restore factors of $\hbar$ and $c$),
\begin{equation}
\label{Amin}
A_{\rm min} = \lambdabar^2 = \left( \frac{\hbar}{m_\nu \, c} \right)^2 \,.
\end{equation}
The detection probability $P$ for a perfectly focused particle
therefore cannot exceed
\begin{equation}
\label{P1}
P = \frac{\sigma}{A^2_{\rm min}} = 
\frac{A_0 \, c^4 \, m_\nu^3}{E_0 \, \hbar^2} \, 
\sqrt{\dfrac{1}{\delta_\nu}} \,.
\end{equation}
If we are to send information reliably, then 
the detection probability should be of order unity. 
Setting $P = 1$ leads to 
\begin{equation}
\label{deltanu1}
\delta_\nu = \frac{A_0^2 \, c^8 \, m_\nu^6}{E_0^2 \hbar^4} \,.
\end{equation}
When traveling at a
speed $c + \delta c$ for a path length $s$, the neutrino acquires a
path length difference of $\delta s$, which compares to its Compton
wavelength $\lambdabar = \hbar/(m_\nu \, c)$ as follows,
\begin{equation}
\label{solve}
\delta s = s \, \frac{\delta c}{c}  = s \, \frac{\delta_\nu}{2} \,.
\end{equation}
The distance traveled by the superluminal neutrino exceeds the  
distance traveled by a light beam by an amount
$\delta s = \lambdabar$ when $s = s_0$ where
\begin{equation}
\label{s0}
s_0 = \frac{2 \, E_0^2 \, \hbar^5}{A_0^2 \, c^9 \, m_\nu^7} 
= 6.63 \times 10^{74} \, {\rm m} \,
\left( \frac{m_\nu}{{\rm eV}/c^2} \right)^{-7} \,.
\end{equation}
Even at a (larger-than-realistic) mass square 
$m_\nu^2 = 1 \, {\rm eV}^2$ for the tachyonic neutrino 
flavor eigenstate, the value of 
$s_0 \sim 10^{74} \, {\rm m} $ far exceeds the commonly
assumed size of the Universe of $10^{26} \, {\rm m}$
by many orders of magnitude.
The permissibility of slightly superluminal propagation
on small length and distance scales has
been discussed in the literature previously (see, e.g.,
Ref.~\cite{AhReSt1998}).  Furthermore, we refer to the experiments in the group
of Nimtz~\cite{EnNi1992,NiSt2008,Ni2009}, which also use a compact apparatus
and rely on the quantum mechanical tunneling effect, which lies outside the
regime of classical mechanics.
Thus, a very slightly superluminal neutrino flavor eigenstate 
with a light mass does not necessarily lead
to a detectable violation of causality.

%
%
\section{Conclusions}

In the current article, we have considered the tachyonic neutrino decay width
against lepton-pair and neutrino-pair 
Cerenkov radiation (LPCR and NPCR, see Fig.~\ref{fig1}),
via the exchange of a virtual $Z^0$ boson.
This process is kinematically allowed for a fast-than-light,
oncoming neutrino.
We use the hypothesis of tachyonic neutrinos described by the tachyonic Dirac 
(not Majorana)
equation~\cite{ChHaKo1985,ChKoPoGa1992,Ko1993,ChKo1994,Ch2012}.
Various kinematic considerations are 
summarized in Sec.~\ref{sec2}.
The tachyonic threshold 
is found according to Eq.~\eqref{Eth_form},
$E_{\thr} \approx 2 \, m_e^2/{m_\nu}$, in the limit
$m_e \gg m_\nu$.
Specificities of the tachyonic decay are studied
in Sec.~\ref{threshLPCR} (threshold calculation), 
Sec.~\ref{threshNPCR} (absence of threshold for NPCR), 
Sec.~\ref{maximum_momentum}
(maximum $q^2$ of the $Z^0$ boson and validity of Fermi theory), 
and Sec.~\ref{whywecannot} (rest frame of the tachyon).
In Sec.~\ref{antipart},
it is shown that, because tachyonic particle states may transform
into antiparticle states upon a Lorentz transformation,
it is indispensable to carry out the calculation 
directly in the lab frame~\cite{Fe1967,Fe1978}.

We continue with a discussion of the interaction Lagrangians
relevant for our studies, from the GWS (Glashow--Weinberg--Salam) model
in Sec.~\ref{int_terms}.
After a brief digression on the degrees of freedom
of three-particle decay processes in Sec.~\ref{degrees}
and a discussion on the general rationale of the 
investigation in Sec.~\ref{rationale},
the calculation of the tachyonic decay width 
is approached in two steps.
In Sec.~\ref{step1}, we first demonstrate that it is 
possible to carry out standard decay rate calculations
of the electroweak theory, directly in the lab frame,
using the muon decay width as an example.
We are finally in the position (see Sec.~\ref{step2})
to carry out the integration of the decay rate,
for the tachyonic dispersion relation,
in the lab frame. 
We find an explicit dependence of the formulas
for the decay rate, $\Gamma$, and the 
energy loss rate, $\dd E/\dd x$, 
on the energy $E_1$ of the incoming neutrino 
(all decay processes are studied within the 
conventions from Fig.~\ref{fig4}).
The main results of our investigations are summarized
in Eqs.~\eqref{GammadEdx1},~\eqref{GammadEdx2} and~\eqref{GammadEdx3};
these formulas describe the decay width of a tachyonic neutrino
against LPCR, and the energy loss per distance of an incoming 
tachyonic neutrino beam. 
This investigation is supplemented, in Sec.~\ref{sec4},
by a calculation of NPCR, culminating in the 
results given in Eq.~\eqref{GammadEdxNPCR} 
for the decay and energy loss rates.

In Sec.~\ref{sec5},
we find that the neutrino pair Cerenkov radiation (NPCR) process,
even if threshold-less, has such a low probability due to the 
weak-interaction physics involved,
that it cannot constrain the tachyonic models,
even for large tachyonic neutrino mass parameters of the 
order of $10^{-2} \, {\rm eV}$.
The lifetime of a tachyonic neutrino against LPCR and NPCR,
assuming a realistic magnitude of the mass parameter,
far exceeds the age of the Universe.
Even a ``Big Bird'' neutrino of energy of $E_\nu \approx 2\,{\rm PeV}$,
would easily survive the travel from the 
blazar PKS B1424-418~(see Ref.~\cite{KaEtAl2016}.
In contrast to Lorentz-violating models, neutrino pair Cerenkov 
radiation does not pressure the
tachyonic neutrino hypothesis.

According to Sec.~\ref{sec5},
we should take the opportunity to clarify that in 
contrast to Ref.~\cite{JeEh2016advhep},
it is actually impossible to 
relate a hypothetical cutoff of the cosmic neutrino 
spectrum at the ``Big Bird'' energy of $2 \, {\rm PeV}$
to the threshold energy for (charged) lepton pair Cerenkov 
radiation, and thus, to a neutrino mass parameter.
The reasons are twofold: First, a further decay process 
exists for tachyonic neutrinos which is not 
subject to a threshold condition, namely
neutrino pair Cerenkov radiation.
Second, the decay and energy loss rates for
{\em both} (charged) lepton as well as 
neutrino pair Cerenkov radiation 
simply are too small to lead to any appreciable energy loss 
for an oncoming tachyonic neutrino flavor eigenstate,
over cosmic distances and time scales.
Formulated differently, we can say that 
that neither lepton nor 
neutrino pair Cerenkov radiation processes 
pressure the tachyonic model in any way.

Finally, we hope that the detailed outline of the
calculation of the decay processes given in Sec.~\ref{sec3}
and~\ref{sec4}k
could be of interest in a wider context,
regarding decay processes and cross sections involving
tachyonic spin-$1/2$ particles.
It is indispensable to introduce further helicity
projectors in the calculation of the bispinor matrix elements
relevant to the process, and the calculations become a little
more complex than for ordinary Dirac spinors
[see Eqs.~\eqref{sumrule1}---\eqref{sumrule3}].
Our approach relies on a consistent formalism developed
for the fundamental tachyonic bispinor solutions,
as reported
in various recent investigations~\cite{JeWu2012epjc,Je2012imag,JeWu2012jpa,JeWu2013isrn,JeWu2014}.

%
%
\section*{Acknowledgments}

Helpful conversation with J. H. Noble and B. J. Wundt are gratefully
acknowledged. This research was supported by the National
Science Foundation (Grant PHY--1403973 and PHY--1710856).
This work was also supported by a J\'anos Bolyai
Research Scholarship of the Hungarian Academy
of Sciences.

\myapp

%
%
\section{Interaction Terms in Electroweak Theory}
\label{appa}

From Eq.~(12.240) of Ref.~\cite{ItZu1980}, we have
for the combined interaction of the left-handed 
and right-handed fermionic currents with the 
$W$ and $Z$ bosons, and the electromagnetic $A$ field,
the following Lagrangian,
\begin{align}
\calL_\ell
=& \;
\bar L_e \, \ii \, \gamma^\mu \, \partial_\mu \; L_e
+ \bar e_R \; \ii \, \gamma^\mu \, \partial_\mu \; e_R
+ \frac{g_w}{\sqrt{2}} \, \left(
\bar \nu_e \, \gamma^\mu \, W^+_\mu \, e_L +
\bar e_L \, \gamma^\mu \, W^-_\mu \, \nu_e \right)
- g_w \, \sin\theta_W \, \bar e \, \gamma^\mu \, A_\mu \, e
\nonumber\\[0.1133ex]
& \;
- \frac{g_w}{2 \cos\theta_W} \,
\bar \nu_e \, \gamma^\mu \, Z_\mu \, \nu_e
+ \frac{g_w}{2} \frac{\cos(2 \theta_W)}{\cos \theta_W} \,
\bar e_L \, \gamma^\mu \, Z_\mu \, e_L -
g_w \, \frac{\sin^2(\theta_W)}{\cos\theta_W} \;
\bar e_R \, \gamma^\mu \, Z_\mu \, e_R \,.
\end{align}
Here, the subscripts $L$ and $R$ denote the left- and right-handed
chirality components, $e$ (as a mathematical symbol,
not subscript) denotes the electron-positron field operator, 
the weak coupling constant is $g_w$,
and $\theta_W$ is the Weinberg angle.
One immediately reads off the electromagnetic Lagrangian 
$\calL_1$ given in Eq.~\eqref{L1}.
Using $e_L = [(1-\gamma^5)/2] e$,
the coupling of the left-handed fermion currents to the 
$W$ boson gives
\begin{align}
\calL_2 =& \;
\frac{g_w}{\sqrt{2}} \, \left(
\bar \nu_e \, \gamma^\mu \, W^+_\mu \, e_L +
\bar e_L \, \gamma^\mu \, W^-_\mu \, \nu_e \right)
= 
\frac{g_w}{2 \sqrt{2}} \, 
\bar e \, \gamma^\mu \, W^-_\mu \, (1-\gamma^5) \nu_e +
{\rm h.c.}
\end{align}
which is just $\calL_2$ [see Eq.~\eqref{L2}].
The coupling term of the neutrino to the $Z$ boson can be 
read off as
\begin{equation}
\label{neutrinoZ}
\calL_3
= - \frac{g_w}{2 \cos\theta_W} \,
\bar \nu_e \, \gamma^\mu \, Z_\mu \, \nu_e
= - \frac{g_w}{4 \cos\theta_W} \,
\bar \nu_e \, \gamma^\mu \, (1 - \gamma^5) \, Z_\mu \, \nu_e \,,
\end{equation}
where we take into account that $\nu_e$ is equal to its 
left-handed chirality component [see Eq.~\eqref{L3}].
The only term which requires a little work is the 
the interaction of the electron current with the $Z$ boson,
\begin{align}
\calL_4 =& \;
\frac{g_w}{2} \frac{\cos(2 \theta_W)}{\cos \theta_W} \,
\bar e_L \, \gamma^\mu \, Z_\mu \, e_L -
g_w \, \frac{\sin^2(\theta_W)}{\cos\theta_W} \;
\bar e_R \, \gamma^\mu \, Z_\mu \, e_R 
\nonumber\\[0.1133ex]
=& \;
\frac{g_w}{2} \frac{1 - 2\sin^2(\theta_W)}{\cos \theta_W} \,
\bar e_L \, \gamma^\mu \, Z_\mu \, e_L -
g_w \, \frac{\sin^2(\theta_W)}{\cos\theta_W} \;
\bar e_R \, \gamma^\mu \, Z_\mu \, e_R
\nonumber\\[0.1133ex]
=& \;
\frac{g_w}{2} \frac{1}{\cos \theta_W} \,
\bar e \, \gamma^\mu \, \frac{1 - \gamma^5}{2} Z_\mu \, e -
g_w \, \frac{\sin^2(\theta_W)}{\cos\theta_W} \;
Z_\mu  \, \left\{ \bar e \, \gamma^\mu \frac{1 - \gamma^5}{2}  \, e +
\bar e \, \gamma^\mu \, \frac{1 + \gamma^5}{2} \, e \right\}
\nonumber\\[0.1133ex]
=& \;
\frac{g_w}{2 \, \cos\theta_W} \,
\overline e \, \left[ \frac12 \, \gamma^\mu (1 - \gamma^5) -
2 \, \sin^2(\theta_W) \, \gamma^\mu \right] \, e \, Z_\mu
\nonumber\\[0.1133ex]
=& \;
- \frac{g_w}{2 \, \cos\theta_W} \,
\overline e \, \left[ 
\left( -\frac12 + 2 \, \sin^2(\theta_W) \right) \, \gamma^\mu +
\frac12 \, \gamma^\mu \, \gamma^5 \right] \, e \, Z_\mu \,.
\nonumber\\[0.1133ex]
=& \;
-\frac{g_w}{2 \, \cos\theta_W} \,
\overline e \, \left[ c_V \, \gamma^\mu
- c_A \, \gamma^\mu \, \gamma^5 \right] \, e \, Z_\mu \,,
\end{align}
where
\begin{equation}
c_V = -\frac12 + 2 \, \sin^2(\theta_W) \,,
\qquad
c_A = -\frac12 \,.
\end{equation}
This result also is in agreement with Eq.~(5.57)
on p.~153 of Ref.~\cite{Ho2002},
up to an overall minus sign which is fixed by the conventions.
According to p.~107 of Ref.~\cite{PDG2012},
the effective Weinberg angle reads as 
\begin{equation}
\sin^2 \theta_W = 0.23146(12) \,,
\qquad
\qquad
\sin^2 \theta_W \approx 0.25 \,,
\end{equation}
which justifies the approximation $c_V \approx 0$, and $c_A \approx -1/2$.
This approximation is often used in the literature
[see the remark preceding Eq.~(2) of Ref.~\cite{CoGl2011}
and p.~153 of Ref.~\cite{Ho2002}].
We also quote from Ref.~\cite{CoGl2011} the $W$ boson mass,
\begin{equation}
M_W = (80.385 \pm 0.015 ) \, \frac{\rm GeV}{c^2}
= 80.385(15) \, \frac{\rm GeV}{c^2} \,,
\end{equation}
and the $Z$ boson mass 
\begin{equation}
M_Z = (91.1876 \pm 0.0021 ) \, \frac{\rm GeV}{c^2}
= 91.1876(21) \, \frac{\rm GeV}{c^2} \,.
\end{equation}
The $W$ and $Z$ masses are 
connected by virtue of the Weinberg angle,
according to Eq.~\eqref{MWMZ}.

%
%
\section{Covariant Pair Production Integrals}
\label{appb}

In our evaluation, for the outgoing electron-positron
pair in the decay of the tachyonic neutrino,
we shall need a few integrals. In the conventions of
Fig.~\ref{fig1}, the outgoing momenta $p_2$ and $p_4$
are on the mass shell, $E_2 = \sqrt{\vec k_2^2 + m_e^2}$ and
$E_4 = \sqrt{\vec k_4^2 + m_e^2}$.
Let us anticipate the results for the integrals $I$, $J$, and $K$,
which are defined as follows,
\begin{subequations}
\label{resIJK}
\begin{align}
\label{resI}
I(q) =& \;
\int \frac{\dd^3 p_2}{2 E_2}
\int \frac{\dd^3 p_4}{ 2 E_4}
\delta^{(4)}( q - p_2 - p_4 ) =
\frac{\pi}{2} \, \sqrt{1 - \frac{4 \, m_e^2}{ q^2 }} \,,
\\[0.133ex]
\label{resJ}
J_{\lambda\rho}(q) =& \;
\int \frac{\dd^3 p_2}{2 E_2}
\int \frac{\dd^3 p_4}{ 2 E_4}
\delta^{(4)}( q - p_2 - p_4 ) \,
\left( p_{2 \lambda} \; p_{4 \rho} \right)  
= 
\sqrt{1 - \frac{4 \, m_e^2}{q^2}} \,
\left[ g_{\lambda \rho} \, \frac{\pi}{24} \,
\left( q^2 - 4 m_e^2 \right) +
q_\lambda \, q_\rho \,
\frac{\pi}{12} \,
\left( 1 + \frac{2 m_e^2}{q^2} \right) \right] \,,
\\[0.133ex]
\label{resK}
K(q) =& \;
\int \frac{\dd^3 p_2}{2 E_2}
\int \frac{\dd^3 p_4}{ 2 E_4}
\delta^{(4)}( q - p_2 - p_4 ) \,
\left( p_2 \cdot p_4 \right)
= \frac{\pi}{4} \,
\sqrt{1 - \frac{4 \, m_e^2}{q^2}} \,
\left( q^2 - 2 m_e^2 \right) \,.
\end{align}
\end{subequations}
By symmetry, one immediately has $J_{\rho\lambda}(q) = J_{\lambda\rho}(q)$.
The evaluation of these integrals is essentially simplified 
because of the Lorentz invariance of the integration measures,
which entails the possibility to choose a coordinate system 
where $q = (q^0, \vec q = \vec 0)$, and then, identify the 
occurrences of $(q^0)^2$ with $q^2$.
The derivation of the results is discussed below.
We observe that because $E_2 = \sqrt{\vec k_2^2 + m_e^2}$
and $E_4 = \sqrt{\vec k_4^2 + m_e^2}$, we have
\begin{align}
\frac{\dd E_4}{\dd |\vec k_4|} 
=& \; \frac{\dd \sqrt{|\vec k|_4^2 + m_e^2}}{\dd |\vec k_4|}
= \frac{\tfrac12 \, 2 |\vec k_4|}{ \sqrt{|\vec k|_4^2 + m_e^2}} 
= \frac{|\vec k_4|}{ E_4 } \,,
\qquad
\qquad
E_4 \, \dd E_4 = |\vec k_4| \, \dd |\vec k_4| \,. 
\end{align}

We shall go through the calculation if the integral $I(q)$ 
in great detail,
\begin{align}
I(q) =& \;
\int \frac{\dd^3 p_2}{2 E_2} \,
\int \frac{\dd^3 p_4}{ 2 E_4} \,
\delta^{(4)}( q - p_2 - p_4 ) \,
= \int \frac{\dd^3 p_2}{2 E_2} \,
\int \frac{\dd^3 p_4}{ 2 E_4} \,
\delta^{(3)}( \vec q - \vec k_2 - \vec k_4 ) \,
\delta( q_0 - E_2 - E_4 ) 
\nonumber\\[0.133ex]
\mathop{=}^{\mbox{$\vec q = \vec 0$}} & \;
\int \frac{\dd^3 p_2}{2 E_2} \,
\int \frac{\dd^3 p_4}{ 2 E_4} \,
\delta^{(3)}( - \vec k_2 - \vec k_4 ) \,
\delta( q_0 - E_2 - E_4 ) 
= \int \frac{\dd^3 p_4}{ 2 E_4} \,
\frac{1}{ 2 E_4} \, \delta( q_0 - 2 E_4 )
\nonumber\\[0.133ex]
=& \;
4 \pi \, \int \frac{\dd E_4 \, E_4 \, p_4}{ 4 E_4^2} \,
\left( \tfrac12 \, \delta( E_4 - \tfrac12 q_0 ) \right)
= \frac{\pi}{2} \, \int \dd E_4 \, 
\sqrt{\frac{E_4^2 - m_e^2}{ E_4^2}} \,
\delta( E_4 - \tfrac12 q_0 ) 
\nonumber\\[0.133ex]
=& \;
\frac{\pi}{2} \, \sqrt{\frac{q_0^2/4 - m_e^2}{ q_0^2/4 }} =
\frac{\pi}{2} \, \sqrt{1 - \frac{4 \, m_e^2}{ q^2 }} \,.
\end{align}
For $J_{\lambda\rho}(q)$, we write
\begin{align}
J_{\lambda\rho}(q) =& \;
\int \frac{\dd^3 p_2}{2 E_2}
\int \frac{\dd^3 p_4}{ 2 E_4}
\delta^{(4)}( q - p_2 - p_4 ) \,
\left( p_{2 \lambda} \; p_{4 \rho} \right) = 
A \, q^2 \, g_{\lambda \rho} + B \, q_\lambda \, q_\rho \,.
\end{align}
Projection onto the tensors $g^{\lambda\rho} $ and 
$q^\lambda \, q^\rho$ leads to
\begin{equation}
\label{eq1}
g^{\lambda\rho} J_{\lambda\rho}(q) =
q^2 \left( 4 A + B \right) \,,
\qquad
q^\lambda \, q^\rho J_{\lambda\rho}(q) =
q^4 ( A + B ) \,.
\end{equation}
Now, we have
\begin{align}
\label{eq2}
g^{\lambda\rho} J_{\lambda\rho}(q) =& \;
\int \frac{\dd^3 p_2}{2 E_2}
\int \frac{\dd^3 p_4}{ 2 E_4} \,
\delta^{(4)}( q - p_2 - p_4 ) \,
\left( p_2 \cdot p_4 \right)
\nonumber\\[0.133ex]
\mathop{=}^{\mbox{$\vec q = \vec 0$}} & \;
\int \frac{\dd^3 p_2}{2 E_2}
\int \frac{\dd^3 p_4}{ 2 E_4} \,
\delta^{(3)}( - \vec k_2 - \vec k_4 ) \,
\delta( q_0 - E_2 - E_4 ) \,
\left( E_2 \, E_4 - \vec k_2 \cdot \vec k_4 \right)
\nonumber\\[0.133ex]
= & \;
\int \frac{\dd^3 p_4}{2 E_4}
\frac{1}{ 2 E_4} \,
\delta( q_0 - 2 E_4 ) \,
\left( E_4^2 + \vec k_4^{\,2} \right)
= \frac{4 \pi}{4} \int \frac{\dd E_4 \, E_4 \, |\vec k_4|}{E_4^2}
\tfrac12 \, \delta( E_4 - \tfrac12 q_0 ) \,
\left( E_4^2 + \vec k_4^{\,2} \right)
\nonumber\\[0.133ex]
= & \;
\pi \int \dd E_4 \,
\sqrt{\frac{E^2_4 - m_e^2}{E_4^2}} \,
\tfrac12 \, \delta( E_4 - \tfrac12 q_0 ) \,
\left( 2 E_4^2 - m_e^2 \right)
= \frac{\pi}{2} \,
\sqrt{\frac{q_0^2/4 - m_e^2}{q_0^2/4}} \,
\left( q_0^2/2 - m_e^2 \right)
\nonumber\\[0.133ex]
= & \; \frac{\pi}{4} \,
\sqrt{1 - \frac{4 \, m_e^2}{q^2}} \,
\left( q^2 - 2 m_e^2 \right) \,.
\end{align}
Yet,
\begin{align}
\label{eq3}
q^\lambda q^\rho J_{\lambda\rho}(q) =& \;
\int \frac{\dd^3 p_2}{2 E_2}
\int \frac{\dd^3 p_4}{ 2 E_4} \,
\delta^{(4)}( q - p_2 - p_4 ) \,
\left( q \cdot p_2 \right)
\left( q \cdot p_4 \right)
\nonumber\\[0.133ex]
\mathop{=}^{\mbox{$\vec q = \vec 0$}} & \;
\int \frac{\dd^3 p_2}{2 E_2}
\int \frac{\dd^3 p_4}{ 2 E_4} \,
\delta^{(3)}( - \vec k_2 - \vec k_4 ) \,
\delta( q_0 - E_2 - E_4 ) \,
\left( q_0 \, E_2 \right)
\left( q_0 \, E_4 \right)
\nonumber\\[0.133ex]
= & \;
\int \frac{\dd^3 p_4}{2 E_4}
\frac{1}{ 2 E_4} \,
\delta( q_0 - 2 E_4 ) \,
\left( q_0 \right)^2 \, E^2_4 
= \frac{4 \pi}{4} \int \frac{\dd E_4 \, E_4 \, |\vec k_4|}{E_4^2}
\tfrac12 \, \delta( E_4 - \tfrac12 q_0 ) \,
\, \left( q_0 \right)^2  \, E_4^2
\nonumber\\[0.133ex]
= & \;
\pi \int \dd E_4 \,
\sqrt{\frac{E^2_4 - m_e^2}{E_4^2}} \,
\delta( E_4 - \tfrac12 q_0 ) \,
\left( q_0 \right)^2  \, E_4^2
= \frac{\pi}{2} \, 
\sqrt{\frac{q_0^2/4 - m_e^2}{q_0^2/4}} \,
\frac{q_0^4}{4} \,
= \frac{\pi}{8} \,
\sqrt{1 - \frac{4 \, m_e^2}{q^2}} \, q^4 \,.
\end{align}
Combining Eqs.~\eqref{eq1},~\eqref{eq2} and~\eqref{eq3}, one 
may finally solve for $A$ and $B$,
\begin{equation}
A =
\frac{\pi}{24 \, q^2} \,
\sqrt{1 - \frac{4 \, m_e^2}{q^2}} \,
\left( q^2 - 4 m_e^2 \right) \,,
\qquad
B =
\frac{\pi}{12 \, q^2} \,
\sqrt{1 - \frac{4 \, m_e^2}{q^2}} \,
\left( q^2 + 2 m_e^2 \right) \,.
\end{equation}
Thus, we confirm the result in Eq.~\eqref{resJ},
\begin{equation}
J_{\lambda \rho}(q) = 
g_{\lambda \rho} \, \frac{\pi}{24} \,
\sqrt{1 - \frac{4 \, m_e^2}{q^2}} \,
\left( q^2 - 4 m_e^2 \right) +
q_\lambda \, q_\rho \,
\frac{\pi}{12} \,
\sqrt{1 - \frac{4 \, m_e^2}{q^2}} \,
\left( 1 + \frac{2 m_e^2}{q^2} \right) \,.
\end{equation}
\color{\grayone}
Finally, contracting with the metric, one has 
\begin{equation}
K(q) = g^{\lambda\rho} J_{\lambda\rho}(q) =
\frac{\pi}{4} \,
\sqrt{1 - \frac{4 \, m_e^2}{q^2}} \,
\left( q^2 - 2 m_e^2 \right) \,,
\end{equation}
confirming the result in Eq.~\eqref{resK}.

\end{document}